\documentclass[aps,pre,showpacs,amsmath,amsfonts,preprintnumbers]{revtex4}
\usepackage{graphicx}% Include figure files
\usepackage{dcolumn}% Align table columns on decimal point
\usepackage{bm}% bold math
\newcommand{\Op}[1]{{\boldsymbol{\mathrm{\hat{#1}}}}}
\newcommand{\beq}{\begin{equation}}
\newcommand{\eeq}{\end{equation}}
\newcommand{\beqar}{\begin{eqnarray}}
\newcommand{\eeqar}{\end{eqnarray}}
\newcommand{\bea}{\begin{eqnarray}}
\newcommand{\eea}{\end{eqnarray}}
\newcommand{\bcen}{\begin{center}}
\newcommand{\ecen}{\end{center}}

\newcommand{\half}{\frac{1}{2}}
\usepackage{amssymb,amsmath}
\usepackage{graphicx}
\begin{document}
\title{Quantum Thermodynamics}

% Authors (add full first names)
\author{Ronnie Kosloff }

% Affiliations / Addresses, add [1] after \address if there is only one affiliation
\address{Institute of Chemistry, The Hebrew University, Jerusalem 91904, Israel}

%Contact information of the corresponding author, add [2] after \corres if there are more than one corresponding author
%\corres{ronnie@fh.huji.ac.il; Tel. 972-26585485 Fax 972-26513742}

% Abstract
\begin{abstract}
Quantum thermodynamics addresses the emergence of thermodynamical laws from quantum mechanics.
The link is based on  the intimate connection of quantum thermodynamics with the theory of open quantum systems.
Quantum mechanics inserts dynamics into thermodynamics giving a sound foundation to finite-time-thermodynamics.
The emergence of the 0-law I-law II-law and III-law of thermodynamics from quantum considerations is presented.
The emphasis is on consistence between the two theories which address the same subject from different foundations.
We claim that inconsistency is the result of faulty analysis pointing to flaws in approximations.
\end{abstract}
% Keywords: add 3 to 10 keywords
\maketitle
%\keyword{Thermodynamics; open quantum systems; heat engines; refrigerators; quantum devices; finite time thermodynamics}

% the fields PACS and MSC may be left empty or commented out if not applicable
%\PACS{}
%\MSC{}

%%%%%%%%%%%%%%%%%%%%%%%%%%%%%%%%%%%%%%%%%%%%%%%%%%%%%%%%%%%%

\section{Introduction}

 {\em Quantum thermodynamics} is the study of thermodynamical processes within the context of quantum dynamics. 
 Thermodynamics preceded  quantum mechanics, consistence with  thermodynamics led to Planck's law,  the dawn of quantum theory. 
Following the ideas of Planck on black body radiation,  Einstein  (1905), quantised the electromagnetic  field \cite{einstein05}.
{\em Quantum thermodynamics} is devoted to unraveling  the intimate connection between the 
laws of thermodynamics and their quantum origin requiring consistency. For many decades the two theories developed separately.
An exception is the study of Scovil et al. \cite{scovil59,geusic67} that showed the equivalence of the Carnot engine \cite{carnot} with the three level Maser,
setting the stage for new developments. 
 
With the establishment of quantum theory the emergence of thermodynamics from quantum mechanics becomes a key issue. The two theories address the same subject from different viewpoints. This requires a consistent view of the state and dynamics of matter. Despite its name, dynamics is absent from most thermodynamic descriptions. The standard theory concentrates on systems close to equilibrium. Quantum mechanics has been used to reintroduce dynamical processes into thermodynamics. In particular, the theory of quantum open systems supplies the framework to separate the system from its environment. 
The Markovian master equation pioneered by Lindblad and Gorini-Kossakowski-Sudarshan (LGKS generator) \cite{lindblad76,kossakowski76} 
is one of the key elements of the theory of quantum thermodynamics \cite{alicki79,k24}.
The dynamical framework allows to reinterpret and justify the theory of finite time thermodynamics \cite{curzon75,berry84,salamon01} which addresses thermodynamical processes taking place in finite time.

A  thermodynamical tradition is learning by example. The model of a heat engine by Carnot \cite{carnot} led to generalisations; the first and second law of thermodynamics.  
A quantum mechanical model of heat engines and refrigerators allows to incorporate dynamics into thermodynamics.
Two types of devices have been studied: reciprocating engines utilising the Otto and Carnot cycle 
and continuous engines resembling lasers and laser cooling devices.  
A reciprocating cycle is partitioned into typically four segments, two {\em adiabats}, where the working system is isolated from the environment,
 and two heat transfer segments either isotherms for the Carnot cycle \cite{k85,lloyd,bender,he02,nori07,guo12} or isochores for the Otto cycle \cite{k116,k152,he09,heJ09,mahler07b,jahnkemahler08,allahmahler08,mahlerbook,k201,k221,abah12,johal11,he12,wang13}. The same cycles were then used as models for refrigerators \cite{k152,k243,k251,mahler07,regal12}.
 
 In quantum thermodynamics {\em adiabats} are modelled by time dependent Hamiltonians. 
 Typically the external control Hamiltonian does not commute with the internal Hamiltonian.
 Infinitely slow operation is the prerequisite for the quantum and thermodynamical adiabatic conditions. 
 Under these conditions, the engine's cycle has zero power.
 To generate finite power the speed of operation has to be increased. Empirically it is known  that faster motion leads to  losses due to friction. 
 The quantum description identifies the source of friction in the inability of the system to stay diagonal in the instantaneous energy frame \cite{k176,k190,k215,k219,k221}.
 Once energy is accounted for, which in an engine cycle occurs on the heat transfer segments, the off-diagonal elements are wiped out. This loss
 leads to the phenomenon of friction \cite{k215,k221}. For special cases when the dynamics can be described by a closed set
 of observables the friction can be reduced or eliminated by requiring that only at the initial and final points of the {\em adiabatic} the state of the systems
 is diagonal in energy. This phenomena has been termed "{\em shortcut to adiabaticity}" \cite{muga09,muga10,muga12,k242,k243,k269}.

In quantum thermodynamics the heat transfer segments are described by a system-bath interaction modelled by open quantum systems techniques.
The LGKS generator \cite{lindblad76,kossakowski76} is typically employed \cite{k85,k116,he02,guo12}. 
For finite power operation the thermal transfer process is never allowed to reach equilibrium with the heat bath which would take an infinite amount of time.
Finally, maximum power output is obtained by optimising
the time allocation on each of the segments of the cycle.
The efficiency of the engine at maximum power can then be compared to the well studied results of finite-time thermodynamics \cite{curzon75,salamon80,berry84,bejan96,salamon01,esposito10,vanderbrook13,wang13b}.
In the limit of high temperature the quantum model converges to the finite-time-thermodynamical result \cite{k85,k87,k221}.

The prime example of a continuous quantum engine is a 3-level laser which has been shown to be limited by Carnot efficiency \cite{scovil59,geusic67}. 
Power optimisation leads to efficiency at maximum power identical to finite time thermodynamics \cite{k24}. 
Many dynamical models have been introduced for different types of continuous quantum engines, all consistent with the laws of thermodynamics 
\cite{alicki79,k24,k122,k156,k169,lloyd,scully03,kieu04,segal06,bushev06,erez08,mahler08,allahmahler08,segal09,he09,esposito09,mahlerbook,popescu10,
scully10,scully11,buttiker12}. 
The prime example of a continuous refrigerator is laser cooling. In this context it is obtained by
reversing the operation of a 3-level laser \cite{k122,k156,k272,k275,ritch12,gieseler12,chen12,zhen12,teufel11,verhagen12}. 
A quantum absorption refrigerator has also been studied
which is a refrigerator with heat as its power source \cite{k169,pekola12,k272,k278}. An example could be a refrigerator driven by sunlight \cite{vandebrock12}.

Amazingly, in all these examples  a thermodynamical description is appropriate up to the level of a single open quantum system
\cite{k24,popescu10,regal12,gershon13}. The common assumption that thermodynamics applies solely to macroscopic systems
is only true in classical mechanics. There are alternative approaches to the emergence of thermodynamical phenomena,
for example based on the complexity of the spectrum of large quantum systems \cite{peres86,deutsch91,serdnicki94,olshani08}. 
For such systems the closed system dynamics of global observables is indistinguishable from dynamics generated by LGKS generators \cite{k237,k256}.

Two contemporary fields of research: Ultra cold matter and  quantum information processing are closely related to quantum thermodynamics.
Cooling mechanical systems unravels their quantum character. As the temperature decreases, degrees of freedom freeze out, leaving a simplified dilute effective Hilbert space. Ultracold quantum systems contributed significantly to our understanding of basic quantum concepts. In addition, such systems form the basis for emerging quantum technologies. The necessity to reach ultralow temperatures requires focus on the cooling process itself, i. e. , quantum refrigeration.  A key  theme in quantum information processes is error correction. The resource for these operations are cold ancilla qubits \cite{shor96}.
It is therefore expected that a quantum computer will be intimately connected to a quantum refrigerator. 

The present review follows the manifestation of the laws of thermodynamics in their quantum dynamical context.
\begin{itemize}
\item{0-law of thermodynamics  deals with the partition of the system from the bath. }
\item{I-law: The first law of thermodynamics is a statement of conservation of energy. }

\item{II-law: The second law is a statement on the irreversibility of dynamics: the breakup of time reversal symmetry.
An  empirical definition: Heat will flow spontaneously from a hot source to a cold sink. These statements
are translated to quantum definitions of positive entropy generation.}

\item{III-law: We will analyse two formulations. The first:
The entropy of any pure substance in thermodynamic equilibrium approaches zero as the temperature approaches the absolute zero.
The second formulation is a dynamical one, known as the unattainability principle:
It is impossible by any procedure, no matter how idealised, to reduce any assembly to absolute zero temperature
in a finite number of operations.}
\end{itemize}

%%%%%%%%%%%%%%%%%%%%%%%%%%%%%%%%%%%%%%%%%%%%%%%%%%%%%%%%%%%%
\section{Quantum open systems}

Quantum thermodynamics is based on a series of idealisations in similar fashion to the ideal gas model which serves classical thermodynamics.
The theory of quantum open systems is  the inspiration for many of these idealisations.
The primary goal of open quantum systems theory
is to develop a {\bf local} dynamical description of the dynamics of a system coupled to an environment termed 
"{\em reduced dynamics}". The variables of the theory are defined by local system observables. These observables constitute
the quantum thermodynamical description.
To account for possible system-bath entanglement the system has to be described by  a mixed state $\Op \rho_S$. 
Observables are obtained from:
$\langle \Op O \rangle =  \mathrm{Tr} \{ \Op \rho_S \Op O \}$ \cite{vNeumann}.

It is customary to assume that the entire world is a large closed system and, therefore, time evolution is governed by
a unitary transformation generated by a global Hamiltonian.
For the combined system bath scenario the  global Hamiltonian can be decomposed into::
\begin{equation}
\label{eq:hamil}
\Op H ~=~ \Op H_S + \Op H_B +\Op H_{SB}~,
\end{equation}
where $\Op H_S$ is the system's Hamiltonian $\Op H_B$ the bath Hamiltonian and $\Op H_{SB}$ the system-bath interaction.
Formally, the state of the system can be obtained from a partial trace over the combined system: 
$\Op \rho_S(t) =  \mathrm{Tr}_B \{ \Op \rho_{SB} (t)\} =  \mathrm{Tr}_B \{ \Op U \Op \rho_{SB} (0) \Op U^{\dagger}\}$, where $\Op U$ is generated by the total Hamiltonian: $\Op U= e^{-\frac{i}{\hbar} \Op H t}$.
Reduced dynamics is an equivalent description utilising only systems operators. The desired outcome is to obtain a local dynamical theory.

There are two major strategies to derive such equations.
The first is based on the weak system-bath coupling expansion which leads to an integro-differential equation for the systems state $\Op \rho_S$ \cite{breuer}:
\begin{equation}
\label{eq:intdif}
\frac{d}{dt} \Op \rho_S(t) = -\frac{i}{\hbar} [\Op H_S , \Op \rho_S (t)] + \int_0^t {\cal K}(t,t') \Op \rho_S (t') dt'~,
\end{equation}
where ${\cal K}$ is the memory kernel,
and an additional assumption that at $t=0$ the system and bath are uncorrelated:
\begin{equation}
\label{eq:corel}
\Op \rho = \Op \rho_S \otimes \Op \rho_B~.
\end{equation}
Assuming  the bath dynamics is fast, Eq. (\ref{eq:intdif}) reduced to a differential form \cite{davis74,davis78}:
\begin{equation}
\frac{d}{dt} \Op \rho_S = -\frac{i}{\hbar} [\Op H_S , \Op \rho_S ] + {\cal L}_D \Op \rho_S~~,
\end{equation}
where ${\cal L}_D$ is the generator of dissipative dynamics.

The alternative approach to reduced dynamics searches for the most general form of the generator of Markovian dynamics
or in more technical terms: A quantum dynamical completely positive semigroup \cite{alicki87,breuer}. 
Kraus has shown \cite{kraus71} that starting from an
uncorrelated initial system and bath state  Eq. (\ref{eq:corel}), a reduced map $\Lambda_S(t) $ to the final time has always the structure:
\begin{equation}
\label{eq:kraus}
\Op \rho_S (t) = \Lambda_S (t) \Op \rho_S (0) = \sum_j \Op K_j \Op \rho_S (0) \Op K_j^{\dagger}~~,
\end{equation}
where $\Op K$ are system operators and $\sum_j \Op K_j  \Op K_j^{\dagger} = \Op I$. When adding a Markovian assumption
$\Lambda_S(t) = e^{{\cal L}t}$ the differential generator ${\cal L}$ of the map becomes \cite{lindblad76,kossakowski76}:
\begin{equation}
\label{eq:lindblad}
\frac{d}{dt} \Op \rho_S = {\cal L} \Op \rho_S= -\frac{i}{\hbar} [\Op H_S , \Op \rho_S ] + 
\sum_j \left( \Op V_j  \Op \rho_S \Op V_j^{\dagger} - \frac{1}{2} \{\Op V_j^{\dagger} \Op V_j, \Op \rho_S \} \right)~,
\end{equation}
where $\Op V$ are system operators and $\Op H_S$ is a renormalised system Hamiltonian. 
Eq. (\ref{eq:lindblad}) is the Lindblad-Gorini-Kossakowski-Sudarshan (LGKS)- semi group generator \cite{lindblad76,kossakowski76}. 
In the LGKS derivation the operators $\Op V_j$ are unrestricted systems operators. Explicit derivations such as in Eq. (\ref{eq:S}) relate
the operators $\Op V_j$ to system-bath coupling terms.
The Markovian dynamics implies also that Eq. (\ref{eq:corel}) is true for all times \cite{lindblad96}.
The completely positive construction assures that the dynamics can be generated from a non-unique Hamiltonian model of some large system.

The Heisenberg representation supplies a direct link to quantum thermodynamical observables. 
The dynamics of an observable represented by the operator $\Op O$ it has the form:
\begin{equation}
\label{eq:hlindblad}
\frac{d}{dt} \Op O = {\cal L}^* \Op O= +\frac{i}{\hbar} [\Op H_S , \Op O] + 
\sum_j \left( \Op V_j  \Op O \Op V_j^{\dagger} - \frac{1}{2} \{\Op V_j \Op V_j^{\dagger}, \Op O \} \right) +\frac{\partial \Op O}{\partial t}~~,
\end{equation}
where  the possibility that the operator $\Op O$ is explicitly time dependent is included.
Eq. (\ref{eq:hlindblad}) allows to follow in time thermodynamical observables, such as energy, for a desired process.

%%%%%%%%%%%%%%%%%%%%%%%%%%%%%%%%%%%%%%%%%%%%%%%%%%%%%%%%%%%%
\section{The 0-Law}

The zero law of thermodynamics is typically stated as:
{\em If A and C are each in thermal equilibrium with B, A is also in equilibrium with C}. 
A thermodynamical description is based on idealised partitions between subsystems. 
An isothermal partition, for example, allows heat to flow
from system to bath maintaining the integrity of the subsystems. Consistency with 
quantum mechanics due to the global structure of the theory is therefore a non-trivial statement.

\subsection{System bath partition}

Quantum thermodynamics idealises that the system can be fully described by local operators,
which is equivalent to the condition:
\begin{equation}
\Op \rho \approx \Op \rho_S \otimes \Op \rho_B~~,
\label{eq:tensor}
\end{equation}
In Eq. (\ref{eq:tensor}) there is no system-bath entanglement, which is also true for  Markovian dynamics.
Thermodynamically the local description of the system is equivalent to the extensivity of its observables.
We conclude that the dynamics represented by the LGKS generator Eq. (\ref{eq:lindblad}) is closely linked to a thermodynamical framework.

\subsection{Thermal equilibrium}

An equilibrium state in general is defined as stationary and stable. This assumption is used to  derive  
the Kubo-Martin-Schwinger stability criterion for thermal equilibrium  \cite{kubo57,schwinger59}.
This criterion will imply that in equilibrium there is no energy current between system and bath.
If we extend the description to a network of baths connected by systems:
\begin{equation}
\Op H ~=~ \Op H_{S_1} + \Op H_{B_a} +\Op H_{S_1B_a} + \Op H_{S_2} + \Op H_{S_1B_a} + \Op H_{B_b}+\Op H_{S_1B_b}+\Op H_{S_2B_a} ...
\label{eq:hmil-net}
\end{equation}
then the KMS condition for a tensor product state of subsystems implies the 0-law for a network of coupled systems.
%%%%%%%%%%%%%%%%%%%%%%%%%%%%%%%%%%%%%%%%%%%%%%%%%%%%%%%%%%%%

\section{The I-law}

The I-law is devoted to the elusive concept of energy conservation.
If the universe is closed and passive, meaning there are no source terms, then its energy is conserved.
This implies that the total evolution is unitary, where the dynamics is generated by a total Hamiltonian $\Op H$, Cf.
Eq. (\ref{eq:hamil}) and Eq. (\ref{eq:hmil-net}). As a result the total energy expectation $\langle \Op H \rangle $ is constant. 

Quantum thermodynamics focusses on the balance of energy of systems coupled to a bath. 
The local change is the sum of the heat currents in and out of the system: Heat flow from the environments and 
power from an external  source. The time derivative of the system's energy  balance becomes:
\begin{equation}
\label{eq:i-law}
\frac{d E_S}{dt }~~=~~\sum_j^N {\cal J}_j +{\cal P}~~,
\end{equation}
where ${\cal J}_j$ is the heat current from the $j$th bath, and ${\cal P}$ is the external power.
The quantum thermodynamic version of the I-law is obtained
by inserting the system Hamiltonian $\Op H_S$ into Eq. (\ref{eq:hlindblad}) leading to:
\begin{equation}
\label{eq:i-law-l}
\frac{d E_S}{dt }~~=~~\langle \frac{\partial \Op H_S}{\partial t} \rangle + \langle {\cal L}_D (\Op H_S) \rangle = \langle \frac{\partial \Op H_S}{\partial t} \rangle+ \sum_j^N {\cal J}_j~~,
\end{equation}
since $[\Op H_S, \Op H_S]=0$ only the dissipative part of ${\cal L}$ appears.
The heat currents ${\cal J}_j$ can be identified as:
\begin{equation}
\label{eq:heat}
{\cal J}_j =  \langle  \Op V_j  \Op H_S \Op V_j^{\dagger} - \frac{1}{2} \{\Op V_j \Op V_j^{\dagger}, \Op H_S \} \rangle
\end{equation}
and the power becomes:
\begin{equation}
\label{eq:power}
{\cal P} = \langle \frac{\partial H_S}{\partial t} \rangle
\end{equation}
Eq. (\ref{eq:heat}) and (\ref{eq:power}) are the dynamical versions of the I-law based on Markovian dynamics \cite{sphon,k23,k24}.

The criticism of relying exclusively on LGKS generators  is that they are not unique. In addition such reliance may violate the II-law. 
The non-uniqueness  is caused  by the substitution  $\Op V \rightarrow \Op V+ i \gamma \Op I$
and $\Op H_S \rightarrow \Op H_S+ \frac{\gamma}{2}(\Op V+ \Op V^{\dagger})$ which maintains  the dynamics generated by Eq. (\ref{eq:hlindblad}) invariant. 
In contradiction, the heat current ${\cal L}_D (\Op H_S)$, is not invariant to this transformation. The origin of this fuzziness of the definition stems
from arbitrariness of accounting for the system bath interaction energy in $\Op H_S$. 
To obtain a consistent definition of the I-law additional restrictions must be imposed.

\subsection{The dynamical generator in the weak system-bath coupling limit}

A unique and consistent approach is obtained by deriving the generator ${\cal L}_D$ in the weak system bath  coupling limit. 
In this limit the interaction energy can be neglected. 
This approach represents  a thermodynamical idealisation: It allowed energy transfer while keeping a tensor product separation between 
the system and bath, i.e. a quantum version of an isothermal partition.

Consider a system and a reservoir (bath), with a "bare" system Hamiltonian $\Op H_S$
and the bath Hamiltonian $\Op H_B$, interacting via the Hamiltonian $\lambda \Op H_{int}=\lambda \Op S\otimes \Op B$.
Here $\lambda $ is
the coupling strength. It is assumed that the bath is stationary:
\begin{equation}
[\Op \rho_B , \Op H_B] = 0,\ \mathrm{Tr}\{ \Op \rho_B\, \Op B \}=0 .
\label{ass}
\end{equation}
The reduced, system-only dynamics in the interaction picture map is defined by  a partial trace over the bath
\begin{equation}
\Op \rho_S (t)=\Lambda (t,0) \Op \rho_S \equiv \mathrm{Tr}_B \left \{ \Op U_{\lambda}(t,0)\Op \rho_S \otimes \Op \rho_B \Op U_{\lambda}(t,0)^{\dagger} \right \}
\label{red_dyn}
\end{equation}
where the unitary propagator in the interaction picture is described by the time ordered exponential
\begin{equation}
\Op U_{\lambda}(t,0) = \mathcal{T}\exp\Bigl\{\frac{-i\lambda}{\hbar}\int_0^t \Op S(s)\otimes \Op B(s)\,ds\Bigr\}
\label{prop_int}
\end{equation}
where
\begin{equation}
\Op S(t) = e^{(i/\hbar)\Op H_S' t} \Op S e^{-(i/\hbar) \Op H_S' t} ,\  \Op B (t)= e^{(i/\hbar)\Op H_B t} \Op B e^{-(i/\hbar)\Op H_B t}.
\label{prop_int1}
\end{equation}
$\Op S(t)$ is defined  with respect to the renormalized, \emph{%
physical}, $\Op H_S'$ and not $\Op H_S$ which can be expressed as
\begin{equation}
\Op H_S'=\Op H_S+\lambda ^{2}\Op H_{1}^{\mathrm{corr}}+\cdots .  
\label{eq:H_S}
\end{equation}
The renormalizing terms containing powers of $\lambda $ are
\emph{ Lamb-shifts} corrections due to the interaction with the bath.  
The lowest order (Born) approximation with respect to the coupling constant $\lambda $ yields 
$\Op H_{1}^{\mathrm{corr}}$. 

A convenient tool to represent the reduced map is  a cumulant expansion 
\begin{equation}
{\Op\Lambda} (t,0)=\exp \sum_{n=1}^{\infty }[\lambda ^{n} {\mathcal K}^{(n)}(t)],
\label{eq:k}
\end{equation}%
One finds that ${\mathcal K}^{(1)}=0$ and the weak coupling limit consists of terminating
the cumulant expansion at $n=2$, hence we denote the generator ${\mathcal K}^{(2)}\equiv {\mathcal K}$: 
\begin{equation}
{\Op \Lambda }(t,0)=\exp [\lambda ^{2} {\mathcal K}(t)+ \Op O(\lambda ^{3})].
\end{equation}%
One obtains%
\begin{equation}
{\mathcal K}(t) \Op \rho_S =\frac{1}{\hbar^2}\int_{0}^{t}ds\int_{0}^{t}du F(s-u) \Op S(s) \Op \rho_S \Op S(u)^{\dag }+(%
\mathrm{similar\ terms})  
\label{eq:K(t)}
\end{equation}%
where $ F(s)= \mathrm{Tr} \{ \Op \rho _{B}\Op B(s) \Op B \}$.  The \emph{similar
terms} in Eq.~(\ref{eq:K(t)}) are of the form $\Op \rho_S \Op S(s)\Op S(u)^{\dagger }$ and $\Op S(s)\Op S(u)^{\dagger } \Op \rho_S $.
\par
The Markov approximation (in the interaction picture) means that for sufficient long time the generator in Eq. (\ref{eq:k}) becomes:
\begin{equation}
{\mathcal K}(t)\simeq t\mathcal{L}  
\label{eq:L}
\end{equation}
where $\mathcal{L}$ is a Linblad-Gorini-Kossakowski-Sudarshan (LGKS) generator \cite{lindblad76,kossakowski76}. 
To find its form   the effective system coupling term $\Op S(t)$ is decomposed into its Fourier components
\begin{equation}
\Op S(t)=\sum_{\{\omega\} } e^{i\omega t}\Op S_{\omega }, ~~ \Op S_{-\omega }= \Op S_{\omega }^{\dagger}
\label{eq:S}
\end{equation}
where the set $\{\omega\}$ contains \emph{Bohr frequencies} of the Hamiltonian 
\begin{equation}
\Op H_S' = \sum_k \epsilon_k |k\rangle\langle k|, ~~  \omega = \epsilon_k - \epsilon_l .  
\label{Bohr}
\end{equation}
Then Eq. (\ref{eq:K(t)}) can be written as
\begin{equation}
{\mathcal K}(t) \Op \rho_S =\frac{1}{\hbar^2}\sum_{\omega ,\omega ^{\prime }} \Op S_{\omega }\Op \rho_S \Op S_{\omega
^{\prime }}^{\dag }\int_{0}^{t}e^{i(\omega -\omega ^{\prime})u}du\int_{-u}^{t-u}  F(\tau )e^{i\omega \tau }d\tau +(\mathrm{similar}~\mathrm{terms})~~, 
\label{eq:K2}
\end{equation}
with the use of two crucial approximations:
\begin{equation}
\int_{0}^{t}e^{i(\omega -\omega ^{\prime })u}du\approx t\delta _{\omega
\omega ^{\prime }}, ~~  \int_{-u}^{t-u}  F(\tau )e^{i\omega \tau }d\tau \approx {G}(\omega)=\int_{-\infty }^{\infty } F(\tau )e^{i\omega \tau }d\tau \geq 0.
\label{eq:rep1}
\end{equation}
This condition works for for $t\gg \max \{1/(\omega -\omega ^{\prime })\}$. These two approximations lead to 
$\mathcal K(t) \Op \rho _{S}=(t/\hbar^2)\sum_{\omega} \Op S_{\omega }\Op \rho _{S}\Op S_{\omega }^{\dag }{G}(\omega )+(\mathrm{similar}$ $%
\mathrm{terms})$, and hence it follows from Eq.~(\ref{eq:L}) that $\mathcal{L}_D$ is a special case of the LGKS generator Eq. (\ref{eq:lindblad}) 
derived for the first time by Davies \cite{davis74}. Returning to the Schr\"odinger picture one obtains the following Markovian master equation: 
\begin{eqnarray}
\frac{d \Op \rho_S }{dt} &=&-\frac{i}{\hbar}[\Op H_S', \Op \rho_S ]+\mathcal{L}_D \Op \rho _S,   \\
\mathcal{L}_D \Op \rho_S  &\equiv &\frac{\lambda ^{2}}{2\hbar^2}\sum_{\{\omega \}}G(\omega
)([\Op S_{\omega },\Op \rho_S \Op S_{\omega }^{\dagger }]+[\Op S_{\omega } \Op \rho_S ,\Op S_{\omega
}^{\dagger }])  
\label{Dav}
\end{eqnarray}
The positivity  $G(\omega )\geq 0$  follows from Bochner's theorem and is a necessary condition for
the complete positivity of the Markovian master equation.

The absence of off-diagonal terms in Eq.~(\ref{Dav}), compared
to Eq.~(\ref{eq:K2}), is the crucial property of the Davies generator which can be 
interpreted as coarse-graining in time of fast oscillating terms. It implies also the commutation of $\mathcal{L}_D$
with the Hamiltonian part $[\Op H_S' ,\bullet ]$. This fixes the ambiguity in 
Eq. (\ref{eq:hlindblad}) of the partition between the Hamiltonian and dissipative parts. 
Markovian behaviour involves a rather complicated cooperation
between system and bath dynamics.  
This means that in phenomenological treatments, \emph{one cannot combine arbitrary system Hamiltonians} $\Op H_S~$%
\emph{with a given LGKS generator}. 
This observation is particularly important in the context of quantum thermodynamics, where  it is tempting to study Markovian
dynamics  with an arbitrary control Hamiltonian. 
Erroneous derivations of the quantum master equation can easily lead to a violation of the laws of thermodynamics cf. next section.
\par
For a bath at thermal equilibrium  the additional Kubo-Martin-Schwinger (KMS) \cite{kubo57,schwinger59} condition holds
\begin{equation}
G(-\omega) = \exp\Bigl(-\frac{\hbar\omega}{k_B T}\Bigr) G(\omega),  
\label{KMS}
\end{equation}
where $T$ is the bath's temperature. As a consequence of (\ref{KMS}) the Gibbs state 
\begin{equation}
\Op \rho_{\beta} = Z^{-1} e^{-\beta \Op H_S'}, \ \beta= \frac{1}{k_B T}  
\label{gibbs atate}
\end{equation}
is a stationary solution of (\ref{Dav}). Under mild conditions (e.g : "the only system operators commuting with $\Op H_S'$ and $\Op S$ are scalars") 
the Gibbs state is a unique stationary state and any initial state relaxes towards equilibrium which is consistent with the "0-th law of thermodynamics". 
The corresponding \emph{thermal generator} in Heisenberg form becomes:
\begin{equation} 
\mathcal{L}_D^* \Op O  = \sum_{\{\omega\geq 0\} }\gamma(\omega)\left ( (\Op S_{\omega },\Op O \Op S_{\omega }^{\dagger }-\frac{1}{2}
\{ \Op S_{\omega }\Op S_{\omega}^{\dagger },\Op O \}) + e^{-\hbar\beta\omega}(\Op S_{\omega }^{\dagger},\Op  O \Op  S_{\omega } 
-\frac{1}{2}\{\Op S_{\omega }^{\dagger } \Op S_{\omega}, \Op O \})\right)
\label{Dav_therm}
\end{equation}
where finally the kinetic coefficients become Fourier transforms of the bath autocorrelation functions:
\begin{equation}
\gamma(\omega)= \frac{\lambda^2}{\hbar^2} \int_{-\infty}^{+\infty} \mathrm{Tr}\left\{ \Op \rho_B \, e^{i \Op H_B t/\hbar}\,\Op B\, e^{-i\Op H_B t/\hbar}\Op B \right\}\, e^{-i \omega t} dt .  
\label{relaxation}
\end{equation}
The weak system bath coupling is the quantum version of the thermodynamic isothermal partition between system and bath.
It preserves the autonomy of the system's observables allowing energy flow across the boundary, thus restoring the definition
of the heat flow ${\cal J}$ Eq. (\ref{eq:heat}) and power Eq. (\ref{eq:power}).

\subsection{Thermal generators for periodic driving fields}

An external perturbation modifying the hamiltonian of the system will also modify
the heat flow. As a result, the LGKS generator has to be renormalised. For a slow change one can adopt the adiabatic approach
and use the temporary systems Hamiltonian to derive ${\cal L}_D$. 
In general, the temporal changes in the system's Hamiltonian have to be incorporated
in the derivation of the dissipative generator. This task is difficult since fast changes may alter the Markovian assumption.
An important class of problems in quantum thermodynamics is periodically driven systems. Periodic heat engines and power driven refrigerators fall into this class.

A derivation of the dissipative generator ${\cal L}_D$ limited to a periodically driven  time dependent Hamiltonian cf. Eq. (\ref{eq:power}) is sketched.

\begin{enumerate}
\item{The system's  renormalised Hamiltonian is now periodic in time:
\begin{equation}
\label{Ham_per}
\Op H_S(t)= \Op H_S (t + \tau),\ \Op U(t,0) \equiv \mathcal{T}\exp\bigl\{-\frac{i}{\hbar}\int_0^t \Op H_S(s)\,ds\bigr\},
\end{equation}
The role of the constant Hamiltonian is played by an effective Hamiltonian $\Op H_{eft}$ defined 
by the spectrum of the periodic propagator:
\begin{equation}
\Op H_{eff} = \sum_k \epsilon_k |k\rangle\langle k| ,\ \Op U(\tau,0)= e^{-i \frac{1}{\hbar}\Op H_{eff} \tau} .
\label{eq:Sq1}
\end{equation}
$\epsilon_k$ are called "quasi-energies".}
\item{The Fourier decomposition (\ref{eq:S}) is replaced by a double Fourier decomposition: 
\begin{equation}
\Op U(t,0)^{\dagger}\,\Op S\, \Op U(t,0)=\sum_{q\in \mathbf{Z}}\sum_{\{\omega\}} e^{i(\omega+ q\Omega) t} \Op S_{\omega q},  
\label{eq:Sq}
\end{equation}
where $\Omega = 2\pi/\tau$  and  $\{\omega\}= \{\epsilon_k - \epsilon_l\}$.
The decomposition  above follows from Floquet theory.}
\item{The generator in the interaction picture is the sum of its Fourier components:
\begin{equation} 
\mathcal{L}  = \sum_{q\in \mathbf{Z}}\sum_{\{\omega\}}=\mathcal{L}_{\omega q}
\label{Dav_per}
\end{equation}
where
\begin{equation} 
\mathcal{L}_{\omega q}\Op \rho_S  = \frac{1}{2}\gamma(\omega+q\Omega)\bigl\{([\Op S_{\omega q}, \Op \rho_S  \Op S_{\omega q}^{\dagger }]+[\Op S_{\omega q}\Op \rho_S ,\Op S_{\omega q}^{\dagger }]) + e^{-\hbar\beta(\omega+q\Omega)}([\Op S_{\omega q}^{\dagger},\Op \rho_S \Op S_{\omega q}] +[\Op S_{\omega q}^{\dagger }\Op \rho_S ,\Op S_{\omega q}])\bigr\} .
\label{Dav_per1}
\end{equation}
}
\end{enumerate}
Returning to the Schr\"odinger picture we obtain the following master equation:
\begin{equation}
{\frac{d\Op \rho_S (t) }{dt}}= -\frac{i}{\hbar}[\Op H_S (t), \Op \rho_S (t)]+\mathcal{L}_D(t)\Op \rho_S (t), ~~ t\geq 0.
  \label{ME_per2}
\end{equation}
where 
\begin{equation}
\mathcal{L}(t)= \mathcal{L}(t+\tau ) =\mathcal{U}(t,0)\mathcal{L}\mathcal{U}(t,0)^{\dagger }, ~~ \mathcal{U}(t,0)\cdot
= U(t,0)\cdot U(t,0)^{\dagger}.
\label{gen_per} 
\end{equation}
In particular, one can represent the solution of (\ref{ME_per2}) in the form
\begin{equation}
\rho(t) = \mathcal{U}(t,0)e^{\mathcal{L}t}\rho(0), ~~ t\geq 0.  
\label{sol}
\end{equation}
Any state, satisfying $\mathcal{L}\tilde{\rho}= 0$, defines a periodic steady state (limit cycle)
\begin{equation}
\tilde{\rho}(t) = \mathcal{U}(t,0)\tilde{\rho} = \tilde{\rho}(t+\tau), ~~ t\geq 0.  
\label{persol}
\end{equation}
Finally, one should notice that in the case of multiple couplings and multiple heat baths the generator $\mathcal{L}$ can be always represented as an appropriate sum of the terms, like Eq. (\ref{Dav_therm}).

\subsection{Heat flows and power for periodically driven open systems}

The heat currents for periodic systems break up to a sum of Fourier components for each bath labeled by index $j$. 
Then the generator in the interaction picture become:
\begin{equation}
\mathcal{L}=\sum_{j=1}^M \sum_{q\in \mathbf{Z}}\sum_{\{\omega\geq 0\}}\mathcal{L}^j_{\omega q},
\label{decomposition}
\end{equation}
where any single $\mathcal{L}^j_{\omega q}$ has a structure of Eq. (\ref{Dav_per1}) with the appropriate $\gamma_j(\omega)$.
The corresponding time-dependent objects satisfy
\begin{equation}
\mathcal{L}^j_{q\omega}(t) \tilde{\rho}^j_{q\omega}(t) = 0, \ \mathcal{L}^j_{q\omega}(t) =\mathcal{U}(t,0)\mathcal{L}^j_{q\omega}\mathcal{U}(t,0)^{\dagger },\   \tilde{\rho}^j_{q\omega}(t) = \mathcal{U}(t,0)\tilde{\rho}^j_{q\omega} = \tilde{\rho}^j_{q\omega}(t+\tau) \  .
\label{inv.state}
\end{equation}
Using the decomposition (\ref{decomposition}), one can define a \emph{local heat current} which corresponds to the exchange of energy $\omega + q\Omega$ with the $j$-th heat bath for any initial state 
\begin{equation}
{\mathcal{J}^j_{q\omega}}(t) = \frac{\omega + q\Omega}{\omega}\mathrm{Tr}\bigl \{( \mathcal{L}^j_{q\omega}(t)\rho(t))\tilde{\bf H}_S(t)\bigr \} ,\ \tilde{\bf H}_S(t)=\mathcal{U}(t,0) \Op H_S ,
\label{curr_loc}
\end{equation}
\par
The heat current associated with the $j$-th bath is a sum of the corresponding local ones
\begin{equation}
{\mathcal{J}^j}(t) = \sum_{q\in \mathbf{Z}}\sum_{\{\omega\geq 0\}}{\mathcal{J}^j_{q\omega}}(t) 
\label{curr_per1}
\end{equation}
where the sum is over multiples of Floquet frequency and Bohr frequencies of the effective Hamiltonian.

We emphasise in this section a dynamical version of the I-law based on examining energy currents between the system and baths.
This derivation for a periodically driven system is also consistent with the II-law of thermodynamics \cite{k114,k122,k278}.

%%%%%%%%%%%%%%%%%%%%%%%%%%%%%%%%%%%%%%%%%%%%%%%%%%%%%%%%%%%%
\section{The II-law}
\label{ii-law}

The second law is a statement on the irreversibility of dynamics or alternatively the breakup of time reversal symmetry.
This should be consistent with the  empirical direct definition: Heat will flow spontaneously from a hot source to a cold sink. 
In classical thermodynamics this statement is equivalent to the statement that the
change in entropy of the universe is positive: $\Delta {\cal S} \ge 0$.
In addition, the entropy generation is additive.
There is considerable confusion in the adaptation of the second-law in quantum mechanics 
where static and dynamical viewpoints are employed. 

From a static viewpoint, for a closed quantum system the II-law of thermodynamics is a consequence of the unitary evolution
\cite{lieb99}. In this approach one accounts for the entropy change before and after a change in the entire system.
A dynamical viewpoint is based on local accounting for the entropy changes in the subsystems and the entropy generated in the baths.

\subsection{Entropy}

In thermodynamics, entropy is related to a concrete process. In quantum mechanics this translates to the ability to measure
and manipulate the system based on the information gathered by measurement \cite{k10}. 
An example is the case of Maxwell's demon which has been resolved by Szilard \cite{szilard29,brillouin,nori09,raizen09}.
There is a close relationship
between the theory of quantum measurement and filtering out an outcome \cite{vNeumann,nielsen}. 
Gathered information can be employed to extract work from a single bath \cite{scully03}. 
This means that the gathered information depends on the particular measurement therefore
for the same system different entropies appear depending on the observable being measured.
Entropy of an observable is associated with the complete projective measurement of an observable $\Op A$
where  the operator $\Op A$ has a spectral decomposition: $\Op A = \sum_i \alpha_i \Op P_i$ and $\Op P_i =| \alpha_i \rangle \langle \alpha_i |$. 
The probability of the outcome is therefore $p_i =tr \{ \Op \rho \Op P_i \}$. The entropy associated with the observable $\Op A$  is the Shannon entropy
with respect to the possible outcomes \cite{k190}:
\begin{equation}
\label{eq:entA}
{\cal S}_A = -\sum_i p_i \ln p_i~~,
\end{equation}
where dimensionless units are chosen for entropy i.e. $k_B=1$.
The most significant observable in thermodynamics is the energy
represented by the Hamiltonian operator $\Op H$ and its associated energy entropy ${\cal S}_E$ \cite{polkovnikov11}.

von Neumann suggested to single out the most informative observable to characterise the entropy of the system.
This invariant  is obtained by minimising the entropy with respect to all possible observables. 
The most informative observable operator commutes with the state of the system $[\Op V, \Op \rho]=0$. 
The entropy of this observable is termed the von Neumann entropy \cite{vNeumann} and is equal to:
\begin{equation}
{\cal S}_{vn} = - tr \{ \Op \rho \ln \Op \rho \}~~~,
\label{eq:entvn}
\end{equation}
As a consequence, ${\cal S}_A \ge {\cal S}_{vn}$ for all observables $\Op A \ne \Op V$. 
${\cal S}_{vn}$ is invariant to a unitary transformation changing the state $\Op \rho$. 
The invariance is a  consequence of the entropy being a functional
of the eigenvalues of $\Op \rho$.  A unitary transformation does not change these eigenvalues.

The von Neumann entropy ${\cal S}_{vn}$ is additive only for a system state which is composed of a tensor product of its subsystems 
$\Op \rho =\prod_j \otimes \Op \rho_j$. 
In the general case the subsystems are entangled. 
If  local measurements are only accessible
then the observable entropy relates to a product of local observables $\Op O =\prod_j \otimes \Op O_j$. 
Since this is a restricted class of operators, the associated entropy
which is the sum of entropies of the subdivision, is always larger than the total von Neumann entropy.
$\sum_j {{\cal S}_{vn}}_j \le {\cal S}_{vn}$. The extreme example is an entangled pure state where for the bipartite case
${\cal S}_{vn}=0$ and ${{\cal S}_{vn}}_1={{\cal S}_{vn}}_2 > 0$. 
This observation is the base for using the reduced state entropy as a measure of entanglement \cite{zurek03,erez05,horodecki07}.
Consider an uncorrelated initial state: Once an interaction 
Hamiltonian turns on, the dynamics will cause the sum of local entropies to increase \cite{k229}. 
A local structure imposed on the total system by thermodynamic partitions is the source of 
local entropy increase. In general, entropy in quantum mechanics is not additive. Once a tensor product partition is imposed
the quantum entropy becomes additive. Alternative sources for entropy increase have been suggested. For example, in a scattering event
the correlation generated by the interaction is lost when the scattering partners reach indefinite distance \cite{k219}. Quantum complexity
can also lead to quantum thermalization \cite{serdnicki94,olshani08} and entropy increase \cite{polkovnikov11}.

\subsection{Quantum networks and quantum devices}

\begin{figure}[htbp]
\center{\includegraphics[height=6cm]{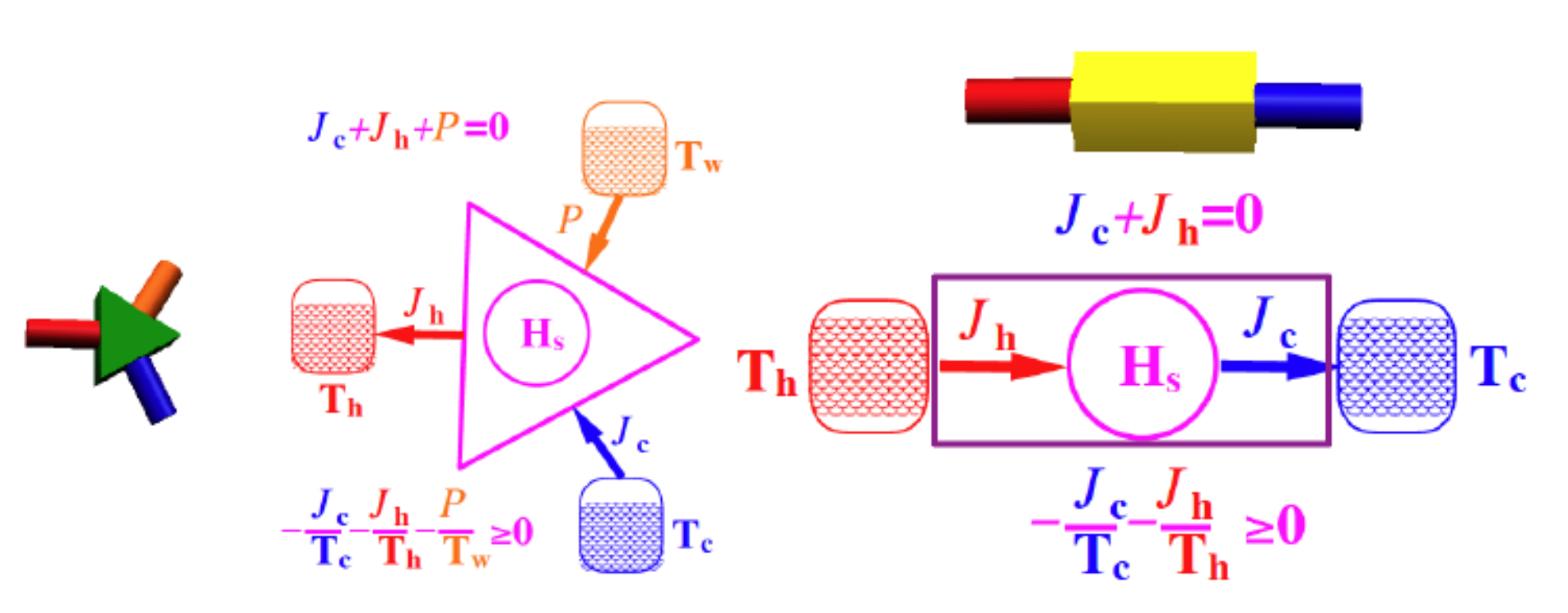}}
\caption{The tricycle on the left and the wire on the right; elementary elements in a quantum network.
The tricycle combines three energy currents.
The tricycle in the figure is connected to three heat baths demonstrating a heat driven refrigerator. 
The wire,  combines two energy currents.
The wire in the figure is connected to a hot and cold bath.
The I-law and II-law are indicated.}
\label{fig:1}
\end{figure}

\begin{figure}[htbp]
\center{\includegraphics[height=8cm]{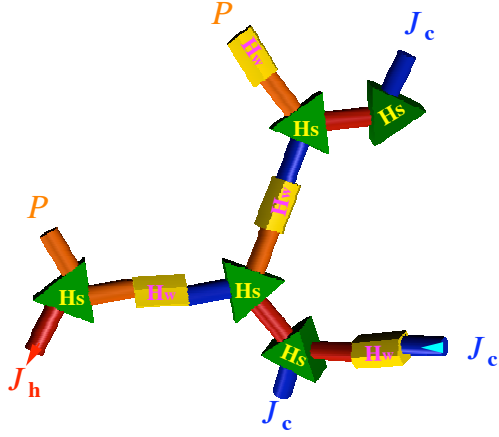}}
\caption{An example of quantum thermodynamical network composed of wires and tricycles.}
\label{fig:2}
\end{figure}

A quantum network is defined as a collection of interconnected quantum systems and baths at different temperatures,
cf. Fig. \ref{fig:1} and \ref{fig:2}.
This network can be decomposed into two elementary segments: A wire and a junction. 
The wire is  a transport line between two segments. In the most simple form it connects two baths.
The junction is a tricycle; a system combining  three currents. A heat engine is a tricycle connected to three baths;
A work bath, and a hot and cold bath.
More complex networks can be constructed from these two basic elements cf. Fig. \ref{fig:2}.
A linear network composed of linearly coupled harmonic oscillators can be decomposed to wires only \cite{paz}.

Quantum networks are subject to the Clausius version of the II-law \cite{clausius1850}:
\begin{itemize}
\item{No process is possible whose sole result is the transfer of heat from a body of lower temperature to a body of higher temperature.}
\end{itemize}
This statement can be generalised to N coupled heat baths in steady state:
\begin{equation}
\label{eq:ii-law}
\sum_j^N \frac{{\cal J}_j}{T_j} ~\le~ 0
\end{equation}

A dynamical version of  the II-law can be proven, based on Spohn's inequality  \cite{spohn78}:
\begin{equation}
\mathrm{Tr}\bigl \{ [\mathcal{L}\rho ][\ln \rho - \ln \tilde{\rho}]\bigr \} \leq 0~,
\label{spohn}
\end{equation}
which is valid for any LGKS generator $\mathcal{L}$ with a stationary state $\tilde{\rho}$.
\par
Computing  the time derivative of the von Neumann entropy  ${\cal  S}_{vn}(t)= -k_B \mathrm{Tr} \{ \Op \rho(t)\ln \Op \rho(t) \}$ 
and applying (\ref{spohn}) one obtains the II-law in the form
\begin{equation}
\frac{d}{dt}{\cal S}_{vn} (t) - \sum_{j=1}^M \frac{\mathcal{J}^j(t)}{T_j} \geq 0~,
\label{IIlaw_periodic}
\end{equation}
where ${\cal  S}_{vn} (t) = - \mathrm{Tr}\bigl \{ \Op \rho(t)\ln \Op \rho(t)\bigr \}$.
\par
The heat currents in the steady state for a periodically driven system $\tilde{\rho}(t)$ are time-independent and given by
\begin{equation}
\tilde{\mathcal{J}^j} = - T_j\sum_{q\in \mathbf{Z}}\sum_{\{\omega\geq 0\}}\mathrm{Tr}\bigl \{( \mathcal{L}^j_{q\omega}\tilde{\rho})\ln \tilde{\rho}^j_{q\omega})\bigr \}.
\label{curr_st}
\end{equation}
In steady-state they satisfy the II-law in the form
\begin{equation}
\sum_{j=1}^M \frac{\tilde{\mathcal{J}}^j}{T_j} \leq 0~,
\label{IIlaw_st}
\end{equation}
while, according to the I-law
\begin{equation}
-\sum_{j=1}^M \tilde{\mathcal{J}^j} = - \tilde{\mathcal{J}} = \bar{\mathcal{P}}~.
\label{power_st}
\end{equation}
is the averaged power (negative when the system acts as a heat engine). In the case of a single heat bath
the heat current is always strictly positive except for the case of no-driving when it is equal to zero.
For the constant Hamiltonian these formulas  are also applicable after removing the index $q$, which implies also that $\sum_{j=1}^M \tilde{\mathcal{J}^j}= 0$. 
In a linear quantum system composed of linearly coupled harmonic oscillators  this statement can be derived directly \cite{paz}.

External power carries with it zero entropy generation. This is also the case of entropy generation originating 
from a bath with infinite temperature $\dot {\cal S} =-{\cal J}/(T \rightarrow \infty)=0$. Also power obtained from pure Gaussian white noise carries with it 
zero entropy generation. Another source of zero entropy is a weak quantum measurement \cite{lajos}. 
It seems counterintuitive but a refrigerator driven by external power, a high temperature bath, 
by white noise or by quantum measurements are all equivalent from a thermodynamical standpoint \cite{k275}.

An illuminating example is of work converted to heat in a driven two-level-system coupled to a bath. 
Work is a zero entropy source, therefore it will generate a current flowing to the bath at any finite temperature.
Consistency with the second law will demand that the work will be dissipated to heat in the bath.
In 1946 Bloch proposed to describe the motion of the components of the macroscopic
nuclear polarisation, ${\cal M}$, subject to an external, 
time-dependent, magnetic field \cite{bloch46}.
The Bloch equations have been extended to the optical regime becoming the workhorse of spectroscopy \cite{eberly75}.
The Bloch equations have been derived and re-derived based on the weak system-bath coupling limit \cite{bloch53}.
Their form has the LGKS structure.
Surprisingly, these sets of equations for strong driving fields violate the II-law \cite{k114}. 
The reason for this violation is that in these derivations the energy levels of the system are not dressed by the external driving field.
A derivation of a generalised master equation within the Floquet theory restores the II-law \cite{k114,alicki2013}.

A similar problem arises for the tricycle case. The three-level laser is such an example.
The equivalence of the three-level laser with the Carnot engine was first derived by Scovil et al. \cite{scovil59,geusic67}.
A dynamical set of equations for the laser known as Lamb's equation \cite{lamb64} also violates the II-law \cite{k122}. 
Again, the remedy is a dressed state framework for deriving the dissipative LGKS generator  \cite{k122,k278}. An interesting 
case is a tricycle driven by a Poissonian noise. Such nose has a unitary component which effectively modifies the 
Hamiltonian of the system. As a result the detailed balance conditions of the hot and cold bath are modified \cite{k272}.
Only this procedure saves the II-law of thermodynamics.

The warning emerging  from these examples is that a careful derivation of the LGKS generator is required in order to be consistent with the II-law
of thermodynamics. Examples of violations have been published \cite{nieuwenhuizen2000} which could be traced to a flaw in the Master equation.

\break

\subsection{Approach to steady state: Limit cycle}

In a quantum network the quantum thermodynamical framework leads to a steady state solution: The limit cycle.
The monotonic approach to steady state can be proven for an evolution generated by    a completely-positive-map $\Lambda$.
\begin{equation}
\Op \rho_f = \Lambda \Op \rho_i~.
\end{equation}
For such a map the conditional entropy:
\begin{equation}
{\cal S}(\Op \rho | \Op \rho_f ) =Tr \{ \Op \rho ( \log \Op \rho -\log \Op \rho_f ) \}~,
\end{equation}
is always decreasing \cite{lindblad74}:
\begin{equation}
{\cal S}(\Lambda \Op \rho | \Lambda \Op \rho_f ) \le {\cal S}(\Op \rho | \Op \rho_f ) ~.
\end{equation}
If $\Op \rho_f$ is an invariant of the map $\Lambda$ then repeated application of the map will lead to a limit $\Op \rho \rightarrow \Op \rho_f$
\cite{frigerio77,frigerio78}.
A quantum network whose generator is a sum of LGKS generators will reach steady state.
A reciprocating quantum engine can be described by a completely positive map  ${\cal U}$ which  is the product of 
the maps of each segment. For example,  a map of the Otto refrigerators ${\cal U}={\cal U}_h{\cal U}_{hc}{\cal U}_c{\cal U}_{ch}$ \cite{k201,k274}.
The above argument means that such cycles will reach a  limit cycle of operation \cite{k201}.

\subsection{The quantum and thermodynamical adiabatic conditions and quantum friction}

Thermodynamical adiabatic processes have no entropy change. Typically, an external control
modifies the state.
A quantum version of an adiabatic process can be modelled by an externally controlled time dependent Hamiltonian ${\Op H}(t)$.
If the system is isolated the dynamics is unitary and therefore ${\cal S}_{vn}$ is a constant. 
For this reason the relevant entropy measure for quantum adiabatic processes is the energy entropy ${\cal S}_E$ cf. Eq. (\ref{eq:entA}).
A quantum adiabatic process is defined by ${\cal S}_E$ constant.
Taking the time derivative leads to:
\begin{equation}
\frac{d {\cal S}_E}{dt} = -\sum_j \dot p_j \log p_j
\end{equation}
where $p_j$ is the expectation of the projection on the instantaneous energy level $\epsilon_j(t)$.
The adiabatic condition is therefore equivalent to no net change in the population of the instantaneous energy levels.
This implies that the Hamiltonian should commute with itself at different times:$[ \Op H(t), \Op H(t')]=0$. 

A generic control Hamiltonian used to modify the system from an initial to a final state 
is typically constructed from a static "drift" Hamiltonian $\Op H_0$
and a time dependent control part $\Op H_C(t)$:
\begin{equation}
\Op H(t) = \Op H_0 + \Op H_C(t)
\end{equation}
Typically, $[ \Op H_0, \Op H_C(t)] \ne 0$ therefore also $[ \Op H(t), \Op H(t')] \ne 0$ and time ordering corrections
mean that strictly adiabatic processes are impossible.
Approximate adiabatic dynamics require therefore a slow change of the control Hamiltonian $\Op H_C(t)$. 
These conditions are defined by the adiabatic parameter 
$\mu= \sum_{ij} |\dot \omega_{ij}/\omega_{ij}^2 | \ll 1 $ where $\omega_{ij} = \hbar (\epsilon_i-\epsilon_j)$ 
are the instantaneous Bohr frequencies \cite{comparat09}. 

The adiabatic condition is an important idealisation in thermodynamics. For example if the initial state is the ground state
then the adiabatic conditions define the path that will require minimum work to reach the final value of the control
which will be the ground state of the modified Hamiltonian. 
This statement can be generalised for an initial thermal state of an harmonic oscillator and a TLS, which will commute with $\Op H(t)$ 
and maintain a thermal state at all times \cite{k243}. 

When  the adiabatic conditions are not fulfilled,  additional work is required to reach the final control value.
For an isolated system this work is recoverable since the dynamics are unitary and can be reversed.
The coherences stored in the off-diagonal elements of the density operator carry the required information to recover the extra energy cost
and reverse the dynamics.
Typically, this energy is not recoverable due to interaction with a bath that causes energy dephasing.
The bath in this case acts like a measuring apparatus of energy.
This lost energy is the quantum version of friction \cite{k176,k190,k221}.
The deviation form adiabatic behaviour can be related to the difference between the energy entropy  and the von Neumann entropy \cite{k274}.

There are several strategies to minimise the effect of quantum friction. 
One possibility, termed {\em quantum lubrication}, is to force the state of the system
to commute with the instantaneous Hamiltonian $[\Op \rho(t) ,\Op H(t)]=0$. 
This can be achieved by adding an external source of phase noise \cite{k215}.
It was found that lubrication could be achieved in a small window of control parameters. 
Outside this window the noise caused additional heating
of the system. In the case of quantum refrigerators this noise was always harmful leading to a minimum temperature the refrigerator can reach \cite{k251,k258}.

Is it possible to find non-adiabatic control solutions with an initial and  final state diagonal in the energy representation
$[\Op \rho_i ,\Op H(0)]=0$, $[\Op \rho_f ,\Op H(t_f)]=0~~$? This possibility, which relies on special dynamical symmetries, 
has been termed {\em shortcut to adiabaticity} \cite{muga09,muga10,muga12,k242,k243,k269,k277}. 
The idea is to optimise the scheduling function $f(t)$ of the control $\Op H_C(t)=\Op V_C f(t)$
in such a way that in the shortest time the frictionless transformation from an initial value of the control function to a final value is achieved.

%%%%%%%%%%%%%%%%%%%%%%%%%%%%%%%%%%%%%%%%%%%%%%%%%%%%%%%%%%%%
\section{The III-law}
\label{sec:iii-law}

Two independent formulations of the III-law of thermodynamics exist, both originally stated by
Nernst \cite{nerst06,nerst06b,nerst18}. The first is a purely static (equilibrium) one, also known as the "Nernst heat theorem":
phrased:
\begin{itemize}
\item{The entropy of any pure substance in thermodynamic equilibrium approaches zero as the temperature approaches
zero.}
\end{itemize}
The second formulation is  dynamical, known as the unattainability principle:
\begin{itemize}
\item{It is impossible by any procedure, no matter how idealised, to reduce any assembly to absolute zero temperature
in a finite number of operations \cite{Fowler39,nerst18}}.
\end{itemize}

There is an ongoing debate on the relations between the two formulations and their relation to the II-law regarding which
and if at all, one of these formulations implies the other \cite{landsberg56,landsberg89,belgiorno03,belgiorno03b}. Quantum considerations 
can illuminate these issues.

At steady state the second law implies that the total entropy production is non-negative, cf. Eq. (\ref{IIlaw_st}).
When the cold bath  approaches the absolute zero temperature,
it is necessary to eliminate the entropy production divergence at the cold side. The entropy production at the cold bath  
when $T_c \rightarrow 0$ scales as 
\begin{equation}
\dot S_c \sim - T_c^{\alpha}~~~,~~~~\alpha \geq 0~~.
\label{eq:III-1}
\end{equation}
For the case when $\alpha=0$ the fulfilment of the second law depends on the entropy production of the other baths, 
which should compensate for the negative entropy production of the cold bath. 
The first formulation of the III-law slightly modifies this restriction. Instead of $\alpha \geq 0$ the III-law imposes $\alpha > 0 $ 
guaranteeing that at  absolute zero the entropy production at the cold bath is zero: $\dot S_c = 0$. 
This requirement leads to the scaling condition of the heat current ${\cal J}_c \sim T_c^{\alpha+1}$.
\par
The second formulation is a dynamical one, known as the unattainability principle;
 \emph{No refrigerator can cool a system to absolute zero temperature at finite time}.
This formulation is more restrictive, imposing limitations on the system bath interaction and the cold bath properties when $T_c \rightarrow 0$ \cite{k275}. 
The rate of temperature decrease of the cooling process should vanish according to the characteristic exponent $\zeta$: 
\begin{equation}
\label{eq:cp}
 \frac{dT_c(t)}{dt} \sim -T_c^{\zeta}, ~~~ T_c\rightarrow 0 ~~.
\end{equation}
 Solving Eq. (\ref{eq:cp}),  leads to; 
 \begin{equation}
 T_c(t)^{1-\zeta}=T_c(0)^{1-\zeta}~-ct~~~~~,~for~\zeta < 1~~,
 \label{eq:iii-law-1} 
 \end{equation}
where $c$ is a positive constant. From Eq. (\ref{eq:iii-law-1})  the cold bath is cooled to zero temperature at finite time for $\zeta < 1$. 
 The III-law requires therefore $\zeta \ge 1$.
In order to evaluate Eq.(\ref{eq:cp}) the heat current can be related to the temperature change:
\begin{equation}
 {\cal J}_c(T_c(t)) = -c_V(T_c(t))\frac{dT_c(t)}{dt}~~.
 \label{25}
\end{equation}
This formulation takes into account the heat capacity  $c_V(T_c)$ of the cold bath. $c_V(T_c)$ is determined by the 
behaviour of the degrees of freedom of the cold bath at low temperature.  Therefore the scaling exponents
can be related $\zeta=1+\alpha - \eta$
where $c_V \sim T_c^{\eta}$ when $T_c \rightarrow 0$.

To get additional insight specific cases are examined. The quantum refrigerator models differ in their operational mode being
either continuous or reciprocating. When $T_c \rightarrow 0$ the refrigerators have to be optimised adjusting to the decreasing temperature.
The receiving mode of the refrigerator has to become occupied to transfer energy. The rate of this process is proportional to a Boltzman
term $~ \omega_c^\gamma \exp[ -\frac{\hbar \omega_c}{k_B T_c} ] $. When optimized for maximum cooling rate
the energy difference of the receiving mode  should scale linearly with temperature $\omega_c \sim T_c$
\cite{k156,k169,k243,k272,k275}. 
Once optimised the cooling power of all refrigerators studied have the same 
dependence on the coupling to the cold bath. This means that the III-law depends on the scaling properties of the heat conductivity $\gamma_c(T_c)$
and the heat capacity $c_V(T_c)$ as $T_c \rightarrow 0$.

\subsection{Harmonic oscillator cold heat bath}

The harmonic heat bath is a generic type of a quantum bath. It includes the electromagnetic field: A photon bath, or a macroscopic piece of solid;
a phonon bath, or Bogliyobov excitations in a Bose Einstein condensate. 
The model assumes  linear coupling of the refrigerator to the bath. 
The standard form of the bath's Hamiltonian is:
\begin{equation}
H_{int} = (b + b^{\dagger})\left(\sum_{k} (g(k)a(k) + \bar{g}(k)a^{\dagger}(k))\right)\ ,\ H_B = \sum_{k}\omega(k)a^{\dagger}(k)a(k)~~,
\label{26}
\end{equation}
where $a(k),a^{\dagger}(k)$ are  annihilation and creation operators for a mode $k$. 
For this model the weak coupling limit procedure leads to the LGKS generator with the cold bath relaxation rate given by
\begin{equation}
\gamma_c \equiv \gamma_c(\omega_c) = \pi(\sum_{k} |g(k)|^2 \delta (\omega(k) - \omega_c)\left[1- e^{-\frac{\hbar \omega(k)}{k_BT_c}}\right]^{-1}~~.
\label{27}
\end{equation}
For the bosonic field in $d$-dimensional space, where $k$ is a wave vector, and with the linear  low-frequency dispersion law ($\omega(k) \sim |k|$)   
the following scaling properties for the cooling rate at low frequencies are obtained
\begin{equation}
\gamma_c \sim  \omega_c^{\kappa}\omega_c^{d-1} \left[1- e^{-\hbar \omega_c/k_BT_c}\right]^{-1} 
\label{28}
\end{equation}
where $\omega_c^{\kappa}$ represents the scaling of the coupling strength $|g(\omega)|^2$ and $\omega_c^{d-1}$ 
the scaling of the density of modes. It implies the following scaling relation for the cold current
\begin{equation}
{\cal J}_c \sim   T_c^{d+\kappa} \Bigl[\frac{\omega_c }{T_c}\Bigr]^{d+\kappa} \frac{1}{e^{\hbar \omega_c/k_B T_c}-1}
\label{29}
\end{equation}
Optimization of Eq. (\ref{29}) with respect to $\omega_c$ leads to the frequency tuning $\omega_c \sim T_c$ and the final current scaling
\begin{equation}
{\cal J}_c^{opt} \sim T_c^{d+\kappa}.
\label{30}
\end{equation}
Taking into account that for low temperatures the heat capacity of the bosonic systems scales like
\begin{equation}
c_{V}(T_c) \sim T_c^d
\label{31}
\end{equation}
which  produces the  scaling of the dynamical equation, Eq. (\ref{eq:cp})
\begin{equation}
\frac{d T_c(t)}{dt} \sim - (T_c)^{\kappa}.
\label{32}
\end{equation}
Similarly, the same scaling Eq. (\ref{32}) is  achieved for the periodically driven refrigerator, with the optimization tuning $\omega_c , \lambda \propto T_c$.\\
The III-law implies a constraint on the form of interaction with a bosonic bath 
\begin{equation}
\kappa \geq 1.
\label{33}
\end{equation}
For standard systems like electromagnetic fields or acoustic phonons with linear dispersion law $\omega(k)= v|k|$ and the formfactor $g(k)\sim |k|/\sqrt{\omega(k)}$ the parameter $\kappa =1$ as for low $\omega$, $|g(\omega)|^2 \sim |k|$. However, the condition (\ref{33}) excludes exotic 
dispersion laws $\omega(k)\sim |k|^{\alpha}$ with $\alpha < 1$ which anyway produce the infinite group velocity forbidden by the relativity theory. Moreover, the popular choice of Ohmic coupling is excluded for systems in dimension $d > 1$. The condition (\ref{33}) can be also compared with the condition 
\begin{equation}
\kappa > 2-d~~~,
\label{34}
\end{equation}
which is necessary to assure the existence of the ground state for the bosonic field interacting by means of the Hamiltonian (\ref{26}).
The third law loses its validity if the cold bath does not have a ground state. For a harmonic bath this could happen if
even one of the effective oscillators has an inverted potential.
\subsection{The existence of a ground state}

A natural physical stability condition which should be satisfied by any model of an open quantum system is that its total Hamiltonian should be bounded from below and should possess a ground state. In the quantum degenerate regime even a mixture of isotopes will segregate and lead to a unique ground state.
In the case of systems coupled linearly to bosonic heat baths it implies the existence of the ground state for the following bosonic Hamiltonian 
(compare with (\ref{26})):
\begin{equation}
H_{bos} = \sum_{k}\bigl\{\omega(k)a^{\dagger}(k)a(k) + (g(k)a(k) + \bar{g}(k)a^{\dagger}(k))\bigr\}~~.
\label{ham_bos}
\end{equation}
Introducing a formal transformation to a new set of bosonic operators
\begin{equation}
 a(k) \mapsto b(k)= a(k) + \frac{\bar{g}(k)}{\omega(k)} ~.
\label{trans}
\end{equation}
we can write
\begin{equation}
H_{bos} = \sum_{k}\omega(k)b^{\dagger}(k)b(k)- E_0, \ E_0 = \sum_{k}\frac{|g(k)|^2}{\omega(k)}
\label{ham_bos1}
\end{equation}
with the formal ground state $|0\rangle$ satisfying
\begin{equation}
b(k)|0\rangle =  0, \ \mathrm{for\ all}\ k . 
\label{ground}
\end{equation}
For the interesting case of an infinite set of modes $\{k\}$, labeled by the $d$-dimensional wave vectors, two problems can appear:

1) The ground state energy $E_0$ can be infinite, i.e. does not satisfy
\begin{equation}
 \sum_{k}\frac{|g(k)|^2}{\omega(k)}< \infty.
\label{vanhove}
\end{equation}
2) The transformation (\ref{trans}) can be implemented by a unitary one, i.e. $b(k) = U a(k) U^{\dagger}$ if and only if
\begin{equation}
 \sum_{k}\frac{|g(k)|^2}{\omega(k)^2}< \infty.
\label{vanhove1}
\end{equation}
Non-existence of such a unitary implies non-existence of the ground state (\ref{ground}) (in the Fock space of the bosonic field) and is called \emph{van Hove phenomenon} \cite{algebricmethods}.
\par
While the divergence of the sums (\ref{vanhove}), (\ref{vanhove1}) (or integrals for infinite volume case) for large $|k|$ can be avoided by putting an \emph{ultra-violet cutoff}, the stronger condition (\ref{vanhove1}) imposes  restrictions on the form of $g(k)$ at low frequencies. Assuming, that $\omega(k)= v|k|$ and $g(k)\equiv g(\omega)$ the condition Eq. (\ref{vanhove1}) is satisfied for the following low-frequency scaling in the $d$-dimensional case
\begin{equation}
|g(\omega)|^2 \sim \omega^{\kappa} ,\ \kappa > 2- d .
\label{vanhove2}
\end{equation}
These conditions on the dispersion relation of the cold bath required for  a ground state are 
identical to the conditions for the III-law Eq. (\ref{34}). The consistency with the III-law ensures the existence of the ground state.

\subsection{Ideal Bose/Fermi gas cold heat bath}
\par
An important generic cold bath consists of a degenerate quantum gas composed of ideal Bose or Fermi gas. The model refrigerator
consists of the working medium of (infinitely) heavy particles with the internal structure approximated (at least at low temperatures) by a 
two-level-system (TLS) immersed in the low density gas at the temperature $T_c$. 
Insight into the III-law comes from realising that the degenerate gas is in equilibrium with a normal part. The external refrigerator
only couples to the normal part. Once the temperature approaches zero the fraction of the normal part decreases, eventually nulling the cooling current.
Another source of excitations are collective excitations of Bogoliubov type \cite{Bogoliubov47}. The low energy tail can be described as a phonon bath
with linear dispersion thus the previous section covered the III-law for these excitations.

The Markovian dynamics of such systems was  derived by Dumcke \cite{dumcke85} in the low density limit and $N$-level internal structure. 
For the case of the TLS there is one receiving Bohr frequency $\omega_c$. 
Cooling occurs due to the non-elastic scattering leading to energy exchange with this frequency \cite{k275}:
\begin{equation}
\label{gamma_ldl}
\gamma_c =2\pi {\bf n } \int d^3\vec{p}\int d^3\vec{p'} \delta(E(\vec{p'}) -E(\vec{p}) - \hbar\omega_c) f_{T_c}(\vec{p}_g)|T(\vec{p'},\vec{p})|^2
\end{equation}
with ${\bf n}$ the particles density, $f_{T_c}(\vec{p}_g)$ the probability distribution of the gas momentum strictly given by Maxwell's distribution, $\vec{p} $ and $\vec{p'} $ are the incoming and outgoing gas particle momentum. $E(\vec{p})= p^2/2m$ denotes the kinetic energy of gas particle.\\   
At low-energies (low-temperature), scattering of neutral gas at 3-d can be characterized by s-wave scattering length $a_s$, having a constant transition matrix, $|T|^2 = (\frac{4\pi a_s}{m})^2$.
For this model the integral (\ref{gamma_ldl}) is calculated
\begin{equation}
\label{gamma_ldl2}
\gamma_c= (4\pi)^4({2\pi m T_c})^{-\half} a_s^2 {\bf n } \omega_c {\cal K}_1(\frac{ \hbar \omega_c}{2 k_B T_c})e^{\frac{\hbar \omega_c}{2 k_B T_c}} ~,
\end{equation}
where ${\cal K}_p(x)$ is the modified Bessel function of the second kind. 
Notice that formula (\ref{gamma_ldl2}) is also valid for an harmonic oscillator instead of TLS, assuming only linear terms in the interaction and using the Born approximation for the scattering matrix.  
\par
Optimizing formula (\ref{25}) with respect to $\omega_c$ leads to $\omega_c\sim T_c$. Then  the scaling of the heat current becomes:
\begin{equation}
{\cal J}_c^{opt} \sim {\bf n} (T_c)^{\frac{3}{2}} ~~.
\end{equation}
When the Bose gas is above the critical temperature for the Bose-Einstein condensation the heat capacity $c_V$ and the density ${\bf n}$ are constants.
Below the critical temperature the density ${\bf n}$ in formula (\ref{gamma_ldl}) should be replaced with the density ${\bf n}_{ex}$ of the exited states, having both 
$c_V, {\bf n}_{ex} $ scale as $ \sim (T_c)^{\frac{3}{2}}$ which finally implies
\begin{equation}
\frac{dT_c(t)}{dt}\sim -(T_c)^{\frac{3}{2}}~~.
\label{gas_scaling}
\end{equation} 
In the case of Fermi gas at low temperatures only the small fraction ${\bf n} \sim T_c$ 
of fermions participate in the scattering process and contribute to the heat capacity, the rest is "frozen" in the "Dirac sea" below the Fermi surface.
Again, this effect modifies in the same way both sides of (\ref{eq:cp}) and therefore (\ref{gas_scaling}) is consistent with the III-law.
Similarly, a possible formation of Cooper pairs below the critical temperature does not influence the scaling (\ref{gas_scaling}).

\begin{figure}[htbp]
\center{\includegraphics[height=7cm]{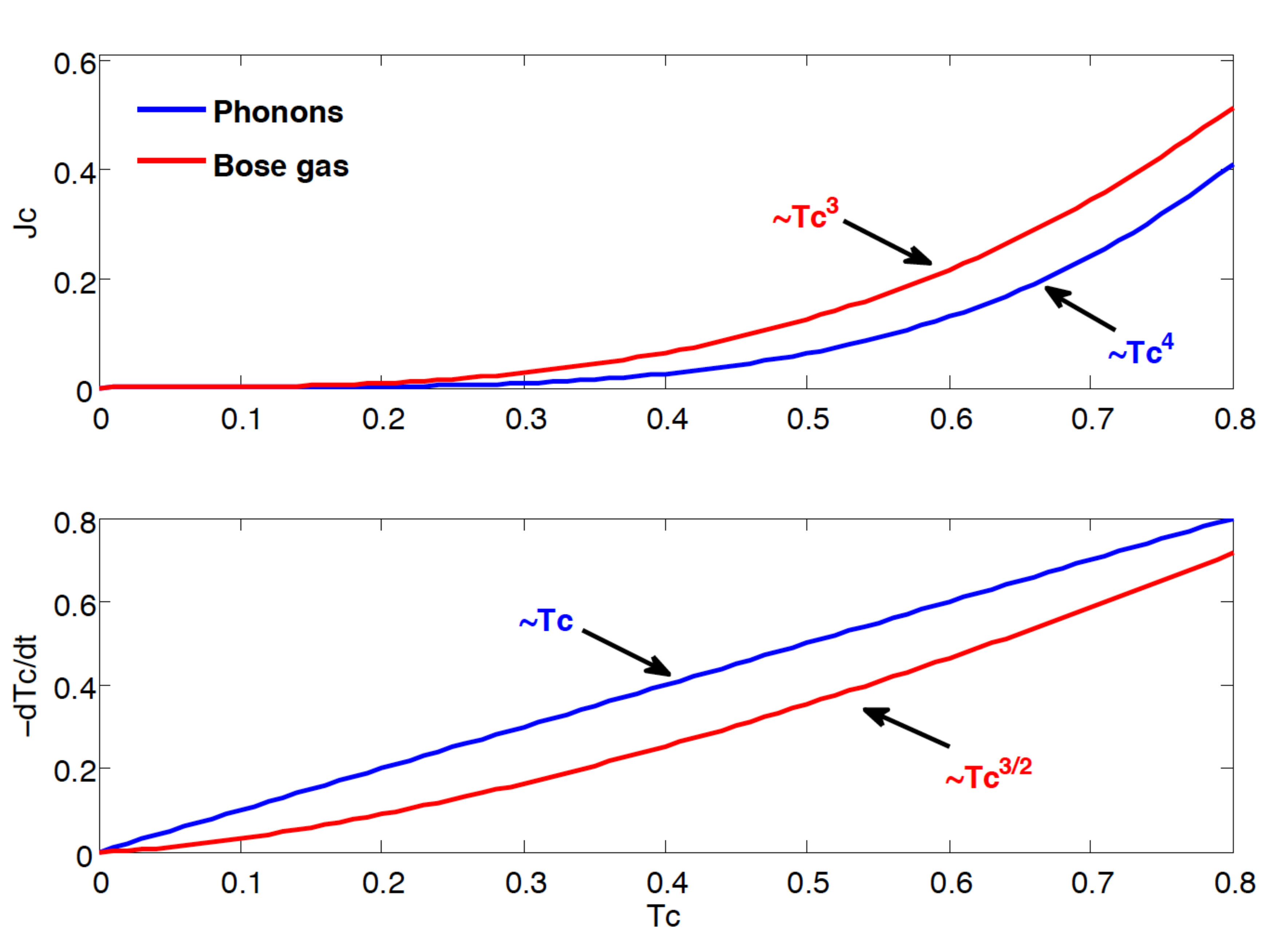}}
\caption{A demonstration of the III-law. The vanishing of the cooling current and the rate of  temperature decrease as $T_c \rightarrow 0$.
The  harmonic bath in 3-D indicated in blue and Bose gas in three dimensions indicated in red. The Bose gas cools faster when $T_c \rightarrow 0$
but its rate of temperature decrease is slower than the harmonic bath.}
\label{fig:3}
\end{figure}

Figure \ref{fig:3} demonstrates the III-law showing the vanishing of the cooling current ${\cal J}_c$ and the temperature decrease rate $\frac{d T_c}{dt}$ as a
function of $T_c$ for the cases of the  harmonic bath and  Bose gas bath. 

\subsection{Thermoelectric refrigerators}
\par
Thermoelectric effect is a combined charge and heat flow between two or more reservoirs \cite{whitney13}.
The non linear interaction that allows such a device to operate is supplied by the coupling between these currents.
The first law is modified by the electrical power ${\cal P}_j^e= V_j I_j$ where $V_j$ is the bias voltage and $I_j$ is the electrical current.The I-law becomes:
\begin{equation}
\label{eq:i-law-e}
\frac{d E_S}{dt }~~=~~\sum_j^N {\cal J}_j +{\cal P}^e_j~~,
\end{equation}
The II-law is not modified  based on scattering theory the charge redistribution does not change entropy \cite{bruneau13}.

The maximum heat current that can be extracted from the cold bath is limited by \cite{whitney13}:
\begin{equation}
{\cal J}_c \le ~ \frac{\pi^2}{6 \hbar ^2}N_c (k_b T_c)^2
\label{eq:thermo-e}
\end{equation}
where $N_c$ is the number of scattering channels. This means that the scaling of the heat current is ${\cal J}_c \sim T_c^2$
and for fermions the heat capacity scales as $c_V \sim T_c$ therefore consistency with the III-law is obtained
with the exponent in Eq. (\ref{eq:cp}) $\zeta=1$.
\par
The dynamical version of the III-law is up for critical analysis \cite{k275}. The examples of quantum refrigerators show
 that the cooling exponents are independent of the type of refrigerator model used. The III-law exponents depend on the cold bath characteristics,
 the ratio between the heat conductivity and the heat capacity for a specific bath. 
 This ratio should scale as $\sim T_c^\zeta$, $\zeta >1$ for the III-law to hold cf, Eq. (\ref{eq:cp}).
There has been a recent challenge to the III-law claiming that zero temperature can be reached \cite{k278,gershon12}.
The present view advocates that this discrepancy is caused by an uncontrolled approximation leading to the particular dispersion used.

%%%%%%%%%%%%%%%%%%%%%%%%%%%%%%%%%%%%%%%%%%%%%%%%%%%%%%%%%%%%
\section{Conclusions}

A dynamical view of quantum thermodynamics was presented. The theory is based on a series of idealisations the main 
one is to impose a local structure through thermodynamic partitions. If the dynamics of the universe is generated
by a hamiltonian, then the conservation of energy and conservation of von Neumann entropy is a trivial statement. 
But without partitions local information on the world cannot be extracted. Our goal is to learn about our local environment.
Quantum thermodynamics follows the laws of thermodynamics within a local structure. 
The theory of open quantum systems is employed to construct such partitions.
The weak coupling limit is an  idealisation employed to construct isothermal partitions which are consistent with the first and second law of thermodynamics.

Quantum thermodynamics is applicable up to the level of a single particle. This means that very simple models have the same 
thermodynamical characteristics of macroscopic devices. For example efficiency at maximum power related to finite-time-thermodynamics. 
Also the quantum and thermodynamical
adiabatic behaviour are closely linked. Deviations lead to friction resulting in reduced efficiency.

The III-law can be thought of as an attempt to isolate completely a subsystem. 
Once a system is cooled to the absolute zero temperature
it reaches a pure ground state and therefore becomes disentangled from the rest of the universe. 
The III-law is a statement that obtaining an isolated pure state is an idealisation impossible at finite time.
 
This review advocates the view that the laws of thermodynamics are true in any quantum circumstance. 
An apparent failure of a quantum model is caused by flaws in the approximation, usually in the derivation 
of the master equations. One can therefore use thermodynamics as a consistency check for approximate quantum theories.

%%%%%%%%%%%%%%%%%%%%%%%%%%%%%%%%%%%%%%%%%%%%%%%%%%%%%%%%%%%%

\section*{Acknowledgements}

I want to thank Eitan Geva, Tova Feldmann, Jose P. Palao, Yair Rezek, Michael Khasin, Amikam Levy, Peter Salamon, Gershon Kurizki, Lajos Diosi and Robert Alicki for sharing their insight on this subject.
Work supported by the Israel Science Foundation. 
Work carried out at the Fritz Haber research center for molecular dynamics,
Hebrew University Jerusalem,
at ITAMP Institute for Theoretical Atomic Molecular and optical Physics, Harvard-Smithsonian, Cambridge MA; and at KITP,
Kalvy Institute for Theoretical Physics, UCSB CA. I thank the hospitality of these institutes.

\break
%==========================================================
%==========================================================
% Back Matter (References and Notes)
%----------------------------------------------------------
% Style and layout of the references
\bibliographystyle{mdpi}
\makeatletter
\renewcommand\@biblabel[1]{#1. }
\makeatother
%----------------------------------------------------------
% Use the following option to include external BibTeX files:
%\bibliography{template}
%----------------------------------------------------------
%\bibliography{dephc1,pub}

\begin{thebibliography}{-------}
\providecommand{\natexlab}[1]{#1}

\bibitem[Einstein(1905)]{einstein05}
Einstein, A.
\newblock "\"Uber einen die Erzeugung und Verwandlung des Lichtes betreffenden
  heuristischen Gesichtspunkt (On a Heuristic Viewpoint Concerning the
  Production and Transformation of Light)".
\newblock {\em Annalen der Physik} {\bf 1905}, {\em 17},~132.

\bibitem[Scovil and du~Bois(1959)]{scovil59}
Scovil, H.E.; du~Bois, E.O.S.
\newblock Three-Level Masers as Heat Engines.
\newblock {\em Phys. Rev. Lett.} {\bf 1959}, {\em 2},~262.

\bibitem[Geusic \em{et~al.}(1967)Geusic, du~Bois, Grasse, and Scovil]{geusic67}
Geusic, J.; du~Bois, E.O.S.; Grasse, R.D.; Scovil, H.E.
\newblock Quantum equivalence of the Carnot Cycle.
\newblock {\em Phys. Rev.} {\bf 1967}, {\em 156},~343.

\bibitem[Carnot(1824)]{carnot}
Carnot, S.
\newblock {\em {R\'eflections sur la Puissance Motrice du Feu et sur les
  Machines propres \`{a} D\'{e}velopper cette Puissance}}; Bachelier: Paris,
  1824.

\bibitem[Lindblad(1976)]{lindblad76}
Lindblad, G.
\newblock On the generators of quantum dynamical semigroups.
\newblock {\em Comm. Math. Phys.} {\bf 1976}, {\em 48},~119.

\bibitem[{Vittorio Gorini Andrzej Kossakowski and E. C. G.
  Sudarshan}(1976)]{kossakowski76}
Gorini, V. ; Kossakowski, A. ;  Sudarshan, E. C. G.
\newblock Completely positive dynamical semigroups of N-level systems.
\newblock {\em J. Math. Phys.} {\bf 1976}, {\em 17},~821.

\bibitem[Alicki(1979)]{alicki79}
Alicki, R.
\newblock Quantum open systems as a model of a heat engine.
\newblock {\em J. Phys A: Math.Gen.} {\bf 1979}, {\em 12},~L103.

\bibitem[Kosloff(1984)]{k24}
Kosloff, R.
\newblock {A Quantum Mechanical Open System as a Model of a Heat Engine.}
\newblock {\em J. Chem. Phys.} {\bf 1984}, {\em 80},~1625--1631.

\bibitem[Curzon and Ahlborn(1975)]{curzon75}
Curzon, F.; Ahlborn, B.
\newblock Efficiency of a Carnot engine at maximum power output.
\newblock {\em Am. J. Phys.} {\bf 1975}, {\em 43},~22.

\bibitem[Bjarne~Andresen and Berry(1984)]{berry84}
Andresen, B.; Salamon, P.; Berry, R.S.
\newblock "Thermodynamics in finite time".
\newblock {\em "Physics Today"} {\bf 1984}, {\em 37:9},~{62}.

\bibitem[{P. Salamon, J.D. Nulton, G. Siragusa, T.R. Andersen and A.
  Limon}(2001)]{salamon01}
Salamon, P.; Nulton, J.D.; Siragusa, G.; Andersen, B.; Limon, A.
\newblock Principles of control thermodynamics.
\newblock {\em Energy} {\bf 2001}, {\em 26},~307.

\bibitem[Geva and Kosloff(1992)]{k85}
Geva, E.; Kosloff, R.
\newblock {A Quantum Mechanical Heat Engine Operating in Finite Time. A Model
  Consisting of Spin $~half~$ Systems as The Working Fluid}.
\newblock {\em J. Chem. Phys.} {\bf 1992}, {\em 96},~3054--3067.

\bibitem[Lloyd(1997)]{lloyd}
Lloyd, S.
\newblock Quantum-mechanical MaxwellÕs demon.
\newblock {\em Phys. Rev. A} {\bf 1997}, {\em 56},~3374.

\bibitem[C.~M.~Bender(2002)]{bender}
Bender, C.~M.; Brody, D. C.; Meister, B.K.
\newblock Entropy and temperature of a quantum Carnot engine.
\newblock {\em Proc. Roy. soc. London, A} {\bf 2002}, {\em 458},~1519.

\bibitem[J.~He and Hua(2002)]{he02}
He, J.;  Chen, J.; Hua, B.
\newblock Quantum refrigeration cycles using spin-$\frac {1}{2}$ systems as
  working substance.
\newblock {\em Phys. Rev. E} {\bf 2002}, {\em 65},~036145.

\bibitem[H.T.~Quan and Nori({2007})]{nori07}
Quan, H.T.; Liu, Y.X.~; Sun, C.P.; Nori, F.
\newblock { Quantum thermodynamic cycles and quantum heat engines }.
\newblock {\em Phys. Rev. E} {\bf {2007}}, {\em 76},~031105.

\bibitem[Guo \em{et~al.}({2012})Guo, Zhang, Su, and Chen]{guo12}
Guo, J.; Zhang, X.; Su, G.; Chen, J.
\newblock {The performance analysis of a micro-/nanoscaled quantum heat
  engine}.
\newblock {\em {Physica-A}} {\bf {2012}}, {\em {391}},~{6432--6439}.

\bibitem[{T. Feldmann, E. Geva, R. Kosloff and P. Salamon}(1996)]{k116}
Feldmann, T.; Geva, E.; Kosloff, R.; Salamon, P.
\newblock {Heat Engines in Finite Time Governed by Master Equations}.
\newblock {\em {Am. J. Phys.}} {\bf 1996}, {\em 64},~485--492.

\bibitem[Feldmann and Kosloff(2000)]{k152}
Feldmann, T.; Kosloff, R.
\newblock {Performance of Discrete Heat Engines and Heat Pumps in Finite Time.}
\newblock {\em Phys. Rev. E} {\bf 2000}, {\em 61},~4774--4790.

\bibitem[H.~Wang and He(2009)]{he09}
Wang, H.; Liu, S.Q.; He, J.Z.
\newblock Thermal entanglement in two-atom cavity QED and the entangled quantum
  Otto engine.
\newblock {\em Phys. Rev. E} {\bf 2009}, {\em 79},~041113.

\bibitem[He~JiZhou and Wei(2009)]{heJ09}
He, J.; Xian, H.; Wei, T.
\newblock "The performance characteristics of an irreversible quantum Otto
  harmonic refrigeration cycle".
\newblock {\em Science in China Series G-Phys. Mech. \& Ast.} {\bf 2009}, {\em
  52},~{1317}.

\bibitem[M.~J.~Henrich(2005)]{mahler07b}
Henrich, M. J.; Rempp, F.; Mahler, G.
\newblock Quantum thermodynamic Otto machines: A spin-system approach.
\newblock {\em Eur. Phys. J.} {\bf 2005}, {\em 151},~157.

\bibitem[T.~Jahnke and Mahler(2008)]{jahnkemahler08}
Jahnke, T. ; Birkov, J; Mahler, G.
\newblock On the nature of thermodynamic extremum principles: The case of
  maximum efficiency and maximum work.
\newblock {\em Ann.Phys.} {\bf 2008}, {\em 17},~88.

\bibitem[A.~E.~Allahverdyan and Mahler(2008)]{allahmahler08}
Allahverdyan, A.~E.; Johal, R.S.; Mahler, G.
\newblock Work extremum principle: Structure and function of quantum heat
  engines.
\newblock {\em Phys. Rev. E} {\bf 2008}, {\em 77},~041118.

\bibitem[J.~Gemmer and Mahler(2009)]{mahlerbook}
Gemmer, J.  Mechel M.; Mahler, G.
\newblock {\em {Quantum Thermodynamics}}; Springer,  2009.

\bibitem[Feldmann and Kosloff(2004)]{k201}
Feldmann, T.; Kosloff, R.
\newblock {Characteristics of the Limit Cycle of a Reciprocating Quantum Heat
  Engine.}
\newblock {\em Phys. Rev. E} {\bf 2004}, {\em 70},~046110.

\bibitem[Yair~Rezek(2006)]{k221}
Rezek, Y.; Kosloff, R.
\newblock {Irreversible performance of a quantum harmonic heat engine}.
\newblock {\em New J. Phys.} {\bf 2006}, {\em 8},~83.

\bibitem[{O. Abah, J. Rossnagel, G. Jacob, S. Deffner, F. Schmidt-Kaler, K.
  Singer and E. Lutz}({2012})]{abah12}
Abah, O.; Rossnagel, J.;  Jacob, G.; Deffner, S.; Schmidt-Kaler, F.; Singer K. ; Lutz, E.
\newblock {Single-Ion Heat Engine at Maximum Power}.
\newblock {\em Phys. Rev. Lett.} {\bf {2012}}, {\em {109}},~{203006)}.

\bibitem[Thomas and Johal(2011)]{johal11}
Thomas, G.; Johal, R.
\newblock Coupled quantum Otto cycle.
\newblock {\em Phys. Rev. E} {\bf 2011}, {\em 83},~031135.

\bibitem[Xian.~He(2012)]{he12}
He, X.; He J.
\newblock Thermal entangled four-level quantum Otto heat engine.
\newblock {\em Sci China-Phys Mech Astron} {\bf 2012}, {\em 55},~1751.

\bibitem[Rui~Wang and Ma(2013)]{wang13}
Wang, R.; Wang, J.; He, J.; Ma, Y.
\newblock { Efficiency at maximum power of a heat engine working with a
  two-level atomic system }.
\newblock {\em Phys. Rev. E} {\bf 2013}, {\em 87},~042119.

\bibitem[{Yair Rezek, Peter Salamon, Karl Heinz Hoffmann and Ronnie
  Kosloff}(2009)]{k243}
Rezek, R.; Salamon, P.; Hoffmann, K. H.; Kosloff, R.
\newblock {The quantum refrigerator: The quest for the absolute zero.}
\newblock {\em Euro. Phys. Lett.} {\bf 2009}, {\em 85},~30008.

\bibitem[Feldmann and Kosloff(2010)]{k251}
Feldmann, T.; Kosloff, R.
\newblock {Minimal temperature of quantum refrigerators}.
\newblock {\em Euro. Phys. Lett.} {\bf 2010}, {\em 89},~20004.

\bibitem[F.~Rempp and Mahler(2007)]{mahler07}
F.~Rempp, M.M.; Mahler, G.
\newblock Cyclic cooling Algorithm.
\newblock {\em Phys. Rev. A} {\bf 2007}, {\em 76},~032325.

\bibitem[A.~M.~Kaufman and Regal({2012})]{regal12}
Kaufman, A. M.; Lester B. J.; Regal, C.A.
\newblock {Cooling a Single Atom in an Optical Tweezer to Its Quantum Ground
  State}.
\newblock {\em Phys. Rev. X} {\bf {2012}}, {\em {2}},~{041014)}.

\bibitem[Kosloff and Feldmann(2002)]{k176}
Kosloff, R.; Feldmann, T.
\newblock {A Discrete Four Stroke Quantum Heat Engine Exploring the Origin of
  Friction.}
\newblock {\em Phys. Rev. E} {\bf 2002}, {\em 65},~055102 1--4.

\bibitem[Feldmann and Kosloff(2003)]{k190}
Feldmann, T.; Kosloff, R.
\newblock {The Quantum Four Stroke Heat Engine: Thermodynamic Observables in a
  Model with Intrinsic Friction}.
\newblock {\em Phys. Rev. E} {\bf 2003}, {\em 68},~016101.

\bibitem[Feldmann and Kosloff(2006)]{k215}
Feldmann, T.; Kosloff, R.
\newblock {Quantum lubrication: Suppression of friction in a first-principles
  four-stroke heat engine.}
\newblock {\em Phys. Rev. E} {\bf 2006}, {\em 73},~025107(R).

\bibitem[{Lajos Di\'osi, Tova Feldmann and Ronnie Kosloff}(2006)]{k219}
Di\'osi, L. ; Feldmann, T. ; Kosloff, R;
\newblock {On exact identity between thermodynamic and informatic entropies in
  a unitary model of friction.}
\newblock {\em International Journal of Quantum Information} {\bf 2006}, {\em
  4},~99--104.

\bibitem[{Xi Chen, A. Ruschhaupt, S. Schmidt, A. del Campo, D. Guery-Odelin, J.
  G. Muga}(2010)]{muga09}
Chen, X. ; Ruschhaupt, A.; Schmidt, S.; del Campo, A.; Guery-Odelin, D.; Muga, J. G.
\newblock Fast optimal frictionless atom cooling in harmonic traps.
\newblock {\em Phys. Rev. Lett.} {\bf 2010}, {\em 104},~063002.

\bibitem[{Xi Chen, I. Lizuain, A. Ruschhaupt, D. GuŽry-Odelin, and J. G.
  Muga}(2010)]{muga10}
Chen, X.; Lizuain, I.; Ruschhaupt, A. ; GuŽry-Odelin, D.;  Muga, J. G.
\newblock Shortcut to Adiabatic Passage in Two- and Three-Level Atoms.
\newblock {\em Phys. Rev. Lett.} {\bf 2010}, {\em 105},~123003.

\bibitem[{S. Ib‡–ez, Xi Chen, E. Torrontegui, J. G. Muga, ; A.
  Ruschhaupt}(2012)]{muga12}
Ib‡–ez, S.; Chen, X.; Torrontegui, E.; Muga, J. G.; Ruschhaupt, A.
\newblock Multiple Schršdinger Pictures and Dynamics in Shortcuts to
  Adiabaticity.
\newblock {\em Phys. Rev. Lett.} {\bf 2012}, {\em 109},~100403.

\bibitem[{Peter Salamon, Karl Heinz Hoffmann, Yair Rezek and Ronnie
  Kosloff}(2009)]{k242}
Salamon, P.; Hoffmann, K. H.; Rezek, Y.   Kosloff, R.
\newblock {Maximum work in minimum time from a conservative quantum system.}
\newblock {\em PCCP} {\bf 2009}, {\em 11},~1027.

\bibitem[K.~H.~Hoffmann and Kosloff(2011)]{k269}
Hoffmann, K. H. ; Salamon, P.; Rezek, Y.; Kosloff, R.
\newblock {Time-optimal controls for frictionless cooling in harmonic traps}.
\newblock {\em Euro. Phys. Lett.} {\bf 2011}, {\em 96},~60015.

\bibitem[P.~Salamon and Berry(1980)]{salamon80}
Salamon, P.;  Nitzan, A.; Andresen, B. ; Berry, R.S.
\newblock Minimum entropy production and the optimization of heat engines.
\newblock {\em Phys. Rev. A} {\bf 1980}, {\em 21},~2115.

\bibitem[Bejan(1996)]{bejan96}
Bejan, A.
\newblock {\em {Entropy Generation Minimization }}; Chemical Rubber Corp,
  1996.

\bibitem[M.~Esposito and den Brook(2010)]{esposito10}
Esposito, M.;  Kawai, R.; Lindenberg, K.; van den Brook, C.
\newblock EfÞciency at Maximum Power of Low-Dissipation Carnot Engines.
\newblock {\em Phys. Rev. Lett.} {\bf 2010}, {\em 105},~150603.

\bibitem[den Brook(2013)]{vanderbrook13}
van den Brook, C.
\newblock Efficiency at maximum power in the low-dissipation limit.
\newblock {\em Eur. Phys. Lett.} {\bf 2013}, {\em 101},~10006.

\bibitem[Wang({2013})]{wang13b}
Wang, H.
\newblock Quantum-mechanical Brayton engine working with a particle in a
  one-dimensional harmonic trap.
\newblock {\em Physica Scripta} {\bf {2013}}, {\em 87},~055009.

\bibitem[Geva and Kosloff(1992)]{k87}
Geva, E.; Kosloff, R.
\newblock {On the Classical Limit of Quantum Thermodynamics in Finite Time}.
\newblock {\em J. Chem. Phys.} {\bf 1992}, {\em 97},~4398--4412.

\bibitem[Geva and Kosloff(1996)]{k122}
Geva, E.; Kosloff, R.
\newblock {The Quantum Heat Engine and Heat Pump: An Irreversible Thermodynamic
  Analysis of The Three-Level Amplifier}.
\newblock {\em J. Chem. Phys.} {\bf 1996}, {\em 104},~7681--7698.

\bibitem[Ronnie~Kosloff and Gordon(2000)]{k156}
Kosloff, R.; Geva, E.; Gordon, J.M.
\newblock {The quantum refrigerator in quest of the absolute zero}.
\newblock {\em J. Appl. Phys.} {\bf 2000}, {\em 87},~8093--8097.

\bibitem[{Jos\'e P. Palao, Ronnie Kosloff, and Jeffrey M. Gordon}(2001)]{k169}
Palao, P. J.;  Kosloff, R,; and Gordon, J.M.
\newblock {Quantum thermodynamic cooling cycle}.
\newblock {\em Phys. Rev. E} {\bf 2001}, {\em 64},~056130--8.

\bibitem[{M. O. Scully, M. S. Zubairy, G. S. Agarwal, H. Walther
  }(2003)]{scully03}
Scully, M. O.; Zubairy, M. S.; Agarwal, G. S.;  Walther, H.
\newblock Extracting work from a single heat bath via vanishing quantum
  coherence.
\newblock {\em Science} {\bf 2003}, {\em 299},~862.

\bibitem[Kieu(2004)]{kieu04}
Kieu, T.D.
\newblock The second law, Maxwell's demon, and work derivable from quantum heat
  engines.
\newblock {\em Phys. Rev. Lett.} {\bf 2004}, {\em 93},~140403.

\bibitem[Segal and Nitzan(2006)]{segal06}
Segal, D.; Nitzan, A.
\newblock Molecular heat pump.
\newblock {\em Phys. Rev. E} {\bf 2006}, {\em 73},~026109.

\bibitem[{"P. Bushev , D. Rotter, A. Wilson, F. Dubin, C. Becher, J. Eschner,
  R. Blatt, V. Steixner, P. Rabl and P. Zoller "}(2006)]{bushev06}
Bushev, P.; Rotter, D.; Wilson, A.; Dubin, F.; Becher, C.;  Eschner, J.; Blatt, R.;
 Steixner, V.;  Rabl, P.;  Zoller, P.
\newblock Feedback Cooling of a Single Trapped Ion.
\newblock {\em Phys. Rev. Lett.} {\bf 2006}, {\em 96},~60010.

\bibitem[Boukobza and Tannor(2008)]{erez08}
Boukobza, E.; Tannor, D.J.
\newblock Thermodynamic analysis of quantum light purification.
\newblock {\em Phys. Rev. A} {\bf 2008}, {\em 78},~013825.

\bibitem[J.~Birjukov(2008)]{mahler08}
Birjukov, J.; Jahnke, T.; Mahler G.
\newblock Quantum thermodynamic processes: a control theory for machine cycles.
\newblock {\em Eur. Phys. J. B} {\bf 2008}, {\em 64},~105.

\bibitem[Segal(2009)]{segal09}
Segal, D.
\newblock Vibrational relaxation in the Kubo oscillator: Stochastic pumping of
  heat.
\newblock {\em J. Chem. Phys.} {\bf 2009}, {\em 130},~134510.

\bibitem[M.~Esposito and den Brook(2009)]{esposito09}
Esposito, M.; Lindberg, k.; van den Brook, C.
\newblock Thermoelectric efficiency at maximum power in a quantum dot.
\newblock {\em Eur. Phys. Lett.} {\bf 2009}, {\em 85},~043003.

\bibitem[N.~Linden(2010)]{popescu10}
Linden, N. ; Popescu, S.; Skrzypczyk, P.
\newblock "How small can thermal machines be? Towards the smallest possible
  refrigerato".
\newblock {\em Phys. Rev. Lett.} {\bf 2010}, {\em 105},~130401.

\bibitem[Scully(2010)]{scully10}
Scully, M.O.
\newblock Quantum Photocell: Using Quantum Coherence to Reduce Radiative
  Recombination and Increase Efficiency.
\newblock {\em Phys. Rev. Lett.} {\bf 2010}, {\em 104},~207701.

\bibitem[{{Anatoly A. Svidzinsky, Konstantin E. Dorfman and Marlan O.
  Scully}}(2011)]{scully11}
Svidzinsky, A. A.; Dorfman K. E.; Scully, M. O.
\newblock Enhancing photovoltaic power by Fano-induced coherence.
\newblock {\em Phys. Rev. A} {\bf 2011}, {\em 84},~053818.

\bibitem[Sothmann and B\"uttiker(2012)]{buttiker12}
Sothmann, B.; B\"uttiker, M.
\newblock Magnon-driven quantum-dot heat engine.
\newblock {\em Eur. Phys. Lett.} {\bf 2012}, {\em 99},~27001.

\bibitem[Levy and Kosloff(2012)]{k272}
Levy, A.; Kosloff, R.
\newblock {Quantum absorption refrigerator}.
\newblock {\em Phys. Rev. Lett.} {\bf 2012}, {\em 108},~070604.

\bibitem[Amikam~Levy and Kosloff(2012)]{k275}
Levy, A.;  Alicki, R.; Kosloff, R.
\newblock {Quantum refrigerators and the third law of thermodynamics}.
\newblock {\em Phys. Rev. E} {\bf 2012}, {\em 85},~061126.

\bibitem[Sandner and Ritsch(2012)]{ritch12}
Sandner, K.; Ritsch, H.
\newblock Temperature Gradient Driven Lasing and Stimulated Cooling.
\newblock {\em Phys. Rev. Lett.} {\bf 2012}, {\em 109},~93601.

\bibitem[Jan~Gieseler and Novotny(2012)]{gieseler12}
Gieseler, J.; Deutsch, B.; Quidant, R.; Novotny, L.
\newblock Subkelvin Parametric Feedback Cooling of a Laser-Trapped
  Nanoparticle.
\newblock {\em Phys. Rev. Lett.} {\bf 2012}, {\em 109},~103603.

\bibitem[{Juncheng Guo, Guozhen Su and Jincan Chen}(2012)]{chen12}
Guo, J.; Su, G.; Chen, J.
\newblock The performance evaluation of a micro/nano-scaled cooler working with
  an ideal Bose gas.
\newblock {\em Physics Letters A} {\bf 2012}, {\em 376},~270.

\bibitem[{Zhen Yi, {Wen-ju} Gu and {Gao-xiang} Li}({2012})]{zhen12}
Yi, Z.; Gu, W-J.; Li, G-X.
\newblock {Sideband cooling of atoms with the help of an auxiliary transition}.
\newblock {\em Phys. Rev. Lett.} {\bf {2012}}, {\em {109}},~{055401}.

\bibitem[{J. D. Teufel, T. Donner, Dale Li, J. W. Harlow, M. S. Allman, K.
  Cicak, A. J. Sirois, J. D. Whittaker, K. W. Lehnert and R. W.
  Simmonds}({2011})]{teufel11}
{J. D. Teufel, T. Donner, Dale Li, J. W. Harlow, M. S. Allman, K. Cicak, A. J.
  Sirois, J. D. Whittaker, K. W. Lehnert and R. W. Simmonds}.
\newblock { Sideband cooling of micromechanical motion to the quantum ground
  state}.
\newblock {\em Nature} {\bf {2011}}, {\em 475},~359.

\bibitem[{E. Verhagen, S. Del\'eglise, S. Weis, A. Schliesser and T. J.
  Kippenberg}({2012})]{verhagen12}
Verhagen, E.; Del\'eglise, S.; Weis, S.; Schliesser A.;  T. J. Kippenberg, T. J.
\newblock {Quantum-coherent coupling of a mechanical oscillator to an optical
  cavity mode}.
\newblock {\em Nature} {\bf {2012}}, {\em 482},~63.

\bibitem[Pekola and Hekking({2007})]{pekola12}
Pekola, J.P.; Hekking, F.W.J.
\newblock {Normal-Metal-Superconductor Tunnel Junction as a Brownian
  Refrigerator}.
\newblock {\em Phys. Rev. Lett.} {\bf {2007}}, {\em {98}},~{210604}.

\bibitem[{Amikam Levy, Robert Alicki and Ronnie Kosloff}(2012)]{k278}
Levy, A.; Alicki, R.; Kosloff, R.
\newblock {Comment on ÒCooling by Heating: Refrigeration Powered by Photons"}.
\newblock {\em Phys. Rev. Lett.} {\bf 2012}, {\em 109},~248901.

\bibitem[B.~Cleuren and den Broeck({2012})]{vandebrock12}
Cleuren, B.R.; van den Broeck, C.
\newblock {Cooling by Heating: Refrigeration Powered by Photons}.
\newblock {\em Phys. Rev. Lett.} {\bf {2012}}, {\em {108}},~{120603}.

\bibitem[D.~Gelbwaser-Klimovsky and Kurizki(2013)]{gershon13}
Gelbwaser-Klimovsky, D.; Alicki, R.; Kurizki, G.
\newblock Minimal universal quantum heat machine.
\newblock {\em Phys. Rev. E} {\bf 2013}, {\em 87},~012140.

\bibitem[Feingold and Peres({1986})]{peres86}
Feingold, M.; Peres, A.
\newblock {Distribution of matrix elements of chaotic systems}.
\newblock {\em Phys. Rev. A} {\bf {1986}}, {\em 34},~591.

\bibitem[Deutch({1991})]{deutsch91}
Deutch, J.M.
\newblock {Quantum statistical mechanics in a closed system}.
\newblock {\em Phys. Rev. A} {\bf {1991}}, {\em 43},~2046.

\bibitem[Srednicki({1994})]{serdnicki94}
Srednicki, M.
\newblock {Chaos and quantum thermalization}.
\newblock {\em Phys. Rev. E} {\bf {1994}}, {\em 50},~888.

\bibitem[Marcos~Rigol and Olshanii({2008})]{olshani08}
Rigol, M.; Dunjko, V; Olshanii, M.
\newblock { Thermalization and its mechanism for generic isolated quantum
  systems}.
\newblock {\em Nature} {\bf {2008}}, {\em 452},~854.

\bibitem[{Michael Khasin and Ronnie Kosloff}(2008)]{k237}
Khasin, M.; Kosloff, R.
\newblock {Efficient simulation of quantum evolution using dynamical coarse
  graining}.
\newblock {\em Phys. Rev. A} {\bf 2008}, {\em 78},~012321.

\bibitem[Khasin and Kosloff(2010)]{k256}
Khasin, M.; Kosloff, R.
\newblock {Algorithm for simulation of quantum many-body dynamics using
  dynamical coarse-graining}.
\newblock {\em Phys. Rev. A} {\bf 2010}, {\em 81},~043635.

\bibitem[Calderbank and Shor({1996})]{shor96}
Calderbank, A.R.; Shor, P.W.
\newblock {Good quantum error-correcting codes exist}.
\newblock {\em Phys. Rev. A} {\bf {1996}}, {\em 54},~1098.

\bibitem[von Neumann(1955)]{vNeumann}
von Neumann, J.
\newblock {\em {Mathematical Foundations of Quantum Mechanics }}; Princeton U.
  P.: Princeton,  1955.

\bibitem[Breuer and Petruccione(2002)]{breuer}
Breuer, H.P.; Petruccione, F.
\newblock {\em Open quantum systems}; Oxford university press,  2002.

\bibitem[Davis({1974})]{davis74}
Davis, E.B.
\newblock {Markovian Master Equations}.
\newblock {\em Comm. Math. Phys.} {\bf {1974}}, {\em {39}},~{91--110}.

\bibitem[Davis({1978})]{davis78}
Davis, E.B.
\newblock {Model of Atomic Radiation}.
\newblock {\em {Annales de Institut Henri Poincare Section A Physique
  Theorique}} {\bf {1978}}, {\em {28}},~{91--110}.

\bibitem[Alicki and Lendi(1987)]{alicki87}
Alicki, R.; Lendi, K.
\newblock {\em {Quantum Dynamical Semigroups and Applications }};
  Springer-Verlag: Berlin,  1987.

\bibitem[Kraus(1971)]{kraus71}
Kraus, K.
\newblock States effects operators.
\newblock {\em Ann.Phys.} {\bf 1971}, {\em 64},~311.

\bibitem[Lindblad({1996})]{lindblad96}
Lindblad, G.
\newblock {On the existence of quantum subdynamics}.
\newblock {\em J. Phys A: Math.Gen.} {\bf {1996}}, {\em {29}},~{4197--4207}.

\bibitem[Kubo({1957})]{kubo57}
Kubo, R.B.
\newblock {Statistical-Mechanical Theory of Irreversible Processes. I. General
  Theory and Simple Applications to Magnetic and Conduction Problems}.
\newblock {\em {J. Phys. Soc. Japan}} {\bf {1957}}, {\em {12}},~{570}.

\bibitem[Martin and Schwinger(1959)]{schwinger59}
Martin, P.C.; Schwinger, J.
\newblock Theory of Many-Particle Systems. I.
\newblock {\em Phys. Rev.} {\bf 1959}, {\em 115},~1342.

\bibitem[Spohn and Lebowitz(1979)]{sphon}
Spohn, H.; Lebowitz, J.
\newblock Irreversible Thermodynamics for Quantum Systems Weakly Coupled to
  Thermal Reservoirs.
\newblock {\em Adv. Chem. Phys.} {\bf 1979}, {\em 38},~109.

\bibitem[{R. Kosloff and M. Ratner}(1984)]{k23}
Kosloff, R.  Ratner, M.A. 
\newblock {Beyond Linear Response: Lineshapes for Coupled Spins or Oscillators
  via Direct Calculation of Dissipated Power.}
\newblock {\em J. Chem. Phys.} {\bf 1984}, {\em 80},~2352--2362.

\bibitem[Eitan~Geva and Skinner(1995)]{k114}
Geva, E.; Kosloff, R.; Skinner, J.
\newblock {On the relaxation of a two-level system driven by a strong
  electromagnetic field}.
\newblock {\em J. Chem. Phys.} {\bf 1995}, {\em 102},~8541--8561.

\bibitem[Lieb and Yngvason(1999)]{lieb99}
Lieb, E.H.; Yngvason, J.
\newblock The Physics and Mathematics of the Second Law of Thermodynamics.
\newblock {\em Phys. Rev.} {\bf 1999}, {\em 310},~1.

\bibitem[{R. Kosloff}(1980)]{k10}
Kosloff, R.
\newblock {Thermodynamic Aspects of the Quantum Measurement Process}.
\newblock {\em Adv. Chem. Phys.} {\bf 1980}, {\em 46},~153--193.

\bibitem[Szilard(1929)]{szilard29}
Szilard, L.
\newblock {On the Minimization of Entropy in a thermodynamic Sytem with
  Interferences of intelligent Beings}.
\newblock {\em Z. Phys.} {\bf 1929}, {\em 53},~840.

\bibitem[Brilluin(1956)]{brillouin}
Brilluin, L.
\newblock {\em {Science and Information Theory }}; Academic Press: New York,
  1956.

\bibitem[Maruyama~Koji({2009})]{nori09}
Maruyama, K.; Nori, F.; Verdal, V.
\newblock { Colloquium: The physics of Maxwell's demon and information }.
\newblock {\em Rev. Mod. Phys.} {\bf {2009}}, {\em 81},~1.

\bibitem[S~Travis~Bannerman and Raizen({2009})]{raizen09}
Travis~Bannerman, S.; Price, G.N.; Viering, K.; Raizen, M.G.
\newblock {Single-photon cooling at the limit of trap dynamics: Maxwell's demon
  near maximum efficiency}.
\newblock {\em New Jour. of Phys.} {\bf {2009}}, {\em 11},~063044.

\bibitem[Nielsen and Chuang(2000)]{nielsen}
Nielsen, M.A.; Chuang, I.L.
\newblock {\em {Quantum Computation and Quantum Information}}; Cambridge
  University Press.: Cambridge,  2000.

\bibitem[Polkovnikov({2011})]{polkovnikov11}
Polkovnikov, A.
\newblock {Microscopic diagonal entropy and its connection to basic
  thermodynamic relations}.
\newblock {\em {ANNALS OF PHYSICS}} {\bf {2011}}, {\em {326}},~{486--499}.

\bibitem[Boukobza and Tannor(2005)]{erez05}
Boukobza, E.; Tannor, D.J.
\newblock Entropy exchange and entanglement in the Jaynes-Cummings model.
\newblock {\em Phys. Rev. A} {\bf 2005}, {\em 78},~063821.

\bibitem[Horodecki \em{et~al.}({2007})Horodecki, Horodecki, Horodecki, and
  Horodecki]{horodecki07}
Horodecki, R.; Horodecki, P.; Horodecki, M.; Horodecki, K.
\newblock {Quantum entanglement}.
\newblock {\em Rev. Mod. Phys.} {\bf {2007}}, {\em {81}},~{865}.

\bibitem[Zurek({2003})]{zurek03}
Zurek, W.H.
\newblock { Quantum discord and MaxwellÕs demons.}
\newblock {\em Phys. Rev. A} {\bf {2003}}, {\em 67},~012320.

\bibitem[{M. Khasin, and R. Kosloff}(2007)]{k229}
Khasin, M.; Kosloff, R.
\newblock {Rise and fall of quantum and classical correlations in open-system
  dynamics}.
\newblock {\em Phys. Rev. A} {\bf 2007}, {\em 76},~012304.

\bibitem[Martinez and Paz(2013)]{paz}
Martinez, E.A.; Paz, J.P.
\newblock Dynamics and thermodynamics of linear quantum open systems.
\newblock {\em Phys. Rev. Lett.} {\bf 2013}, {\em 110},~130406.

\bibitem[Clausius(1850)]{clausius1850}
Clausius, R.
\newblock "Ueber Die Bewegende Kraft Der WŠrme Und Die Gesetze, Welche Sich
  Daraus FŸr Die WŠrmelehre Selbst Ableiten Lassen".
\newblock {\em Annalen der Physik} {\bf 1850}, {\em 79},~368.

\bibitem[Spohn and Lebowitz(1978)]{spohn78}
Spohn, H.; Lebowitz, J.
\newblock Irreversible Thermodynamics for Quantum Systems Weakly Coupled to
  Thermal Reservoirs.
\newblock {\em Adv. Chem. Phys.} {\bf 1978}, {\em 109},~38.

\bibitem[Diosi(2008)]{lajos}
Diosi, L.
\newblock Non-Markovian continuous quantum measurement of retarded observables.
\newblock {\em Phys. Rev. Lett.} {\bf 2008}, {\em 100},~080401.

\bibitem[Bloch({1946})]{bloch46}
Bloch, F.
\newblock {Nuclear Induction}.
\newblock {\em Phys. Rev.} {\bf {1946}}, {\em 70},~460.

\bibitem[Allen and Eberly(1975)]{eberly75}
Allen, L.C.; Eberly, J.H.
\newblock {\em {Optical Resonance and Two-Level Atoms }}; Wiley, Courier Dover
  Publications: New York,  1975.

\bibitem[Wangsness and Bloch({1953})]{bloch53}
Wangsness, R.K.; Bloch, F.
\newblock {The Dynamical Theory of Nuclear Induction}.
\newblock {\em Phys. Rev.} {\bf {1953}}, {\em 89},~729.

\bibitem[{Krzysztof Szczygielski, David Gelbwaser-Klimovsky and Robert
  Alicki}({2013})]{alicki2013}
Szczygielski, K.; Gelbwaser-Klimovsky, D.; Alicki, R.
\newblock {Markovian master equation and thermodynamics of a two-level system
  in a strong laser field}.
\newblock {\em Phys. Rev. E} {\bf {2013}}, {\em 87},~012120.

\bibitem[Lamb({1964})]{lamb64}
Lamb, W.
\newblock {Theory of an Optical Maser}.
\newblock {\em Phys. Rev.} {\bf {1964}}, {\em 134},~1429.

\bibitem[Allahverdyan and Nieuwenhuizen(2000)]{nieuwenhuizen2000}
Allahverdyan, A.E.; Nieuwenhuizen, T.M.
\newblock Extraction of Work from a Single Thermal Bath in the Quantum Regime.
\newblock {\em Phys. Rev. Lett.} {\bf 2000}, {\em 85},~1799.

\bibitem[Lindblad({1974})]{lindblad74}
Lindblad, G.
\newblock {Expectation and entropy inequalities for finite quantum systems}.
\newblock {\em Comm. Math. Phys.} {\bf {1974}}, {\em {39}},~{111--119}.

\bibitem[Frigerio(1977)]{frigerio77}
Frigerio, A.
\newblock Quantum dynamical semigroups and approach to equilibrium.
\newblock {\em Lett. Math. Phys.} {\bf 1977}, {\em 2},~79.

\bibitem[Frigerio(1978)]{frigerio78}
Frigerio, A.
\newblock Stationary states of quantum dynamical semigroups.
\newblock {\em Comm. Math. Phys.} {\bf 1978}, {\em 63},~269.

\bibitem[Feldmann and Kosloff(2012)]{k274}
Feldmann, T.; Kosloff, R.
\newblock {Short time cycles of purely quantum refrigerators}.
\newblock {\em Phys. Rev. E} {\bf 2012}, {\em 85},~051114.

\bibitem[Comparat({2009})]{comparat09}
Comparat, D.
\newblock General conditions for quantum adiabatic evolution.
\newblock {\em Phys. Rev. A} {\bf {2009}}, {\em 80},~012106.

\bibitem[Kosloff and Feldmann(2010)]{k258}
Kosloff, R.; Feldmann, T.
\newblock {Optimal performance of reciprocating demagnetization quantum
  refrigerators}.
\newblock {\em Phys. Rev. E} {\bf 2010}, {\em 82},~011134.

\bibitem[F.~Boldt and Kosloff(2012)]{k277}
Boldt, F.; Hoffmann, K. H.; Salamon, P.; Kosloff, R.
\newblock {Time-optimal processes for interacting spin systems}.
\newblock {\em Euro. Phys. Lett.} {\bf 2012}, {\em 99},~40002.

\bibitem[Nernst(1906{\natexlab{a}})]{nerst06}
Nernst, W.
\newblock {Ueber die Berechnung chemischer Gleichgewichte aus thermischen
  Messungen}.
\newblock {\em {Nachr. Kgl. Ges. Wiss. G\"ott.}} {\bf 1906}, {\em 1},~40.

\bibitem[Nernst(1906{\natexlab{b}})]{nerst06b}
Nernst, W.
\newblock {U\"ber die Beziehung zwischen Wa\"rmeentwicklung und maximaler
  Arbeit bei kondensierten Systemen}.
\newblock {\em {er. Kgl. Pr. Akad. Wiss.}} {\bf 1906}, {\em 52},~933.

\bibitem[Nernst(1918)]{nerst18}
Nernst, W.
\newblock {\em {The theoretical and experimental bases of the New Heat Theorem
  Ger., Die theoretischen und experimentellen Grundlagen des neuen
  Wa\"rmesatzes}}; W. Knapp: Halle,  1918.

\bibitem[Fowler and Guggenheim(1939)]{Fowler39}
Fowler, R.H.; Guggenheim, E.A.
\newblock {\em {Statistical Thermodynamics}}; Cambridge university press,
  1939.

\bibitem[Landsberg(1956)]{landsberg56}
Landsberg, P.T.
\newblock Foundations of Thermodynamics.
\newblock {\em Rev. Mod. Phys.} {\bf 1956}, {\em 28},~363.

\bibitem[Landsberg(1989)]{landsberg89}
Landsberg, P.T.
\newblock A comment on Nernst's theorem.
\newblock {\em J. Phys A: Math.Gen.} {\bf 1989}, {\em 22},~139.

\bibitem[Belgiorno(2003{\natexlab{a}})]{belgiorno03}
Belgiorno, F.
\newblock Notes on the third law of thermodynamics I.
\newblock {\em J. Phys A: Math.Gen.} {\bf 2003}, {\em 36},~8165.

\bibitem[Belgiorno(2003{\natexlab{b}})]{belgiorno03b}
Belgiorno, F.
\newblock Notes on the third law of thermodynamics II.
\newblock {\em J. Phys A: Math.Gen.} {\bf 2003}, {\em 36},~8195.

\bibitem[Emch(1972)]{algebricmethods}
Emch, G.G.
\newblock {\em {Algebraic Methods in Statistical Mechanics and Quantum Field
  Theory }}; Wiley Interscience: New York,  1972.

\bibitem[Bogoliubov({1947})]{Bogoliubov47}
Bogoliubov, N.N.
\newblock {On the theory of superfluidity}.
\newblock {\em J. Phys. (USSR)} {\bf {1947}}, {\em 11},~23.

\bibitem[Dumcke(1985)]{dumcke85}
Dumcke, R.
\newblock {The low density limit for an N-level system interacting with a free
  Bose or Fermi gas }.
\newblock {\em Comm. Math. Phys.} {\bf 1985}, {\em 97},~331.

\bibitem[Whitney({2013})]{whitney13}
Whitney, R.S.
\newblock {Thermodynamic and quantum bounds on nonlinear dc thermoelectric
  transport}.
\newblock {\em Phys. Rev. B} {\bf {2013}}, {\em {87}}.

\bibitem[L.~Bruneau and Pillet({2013})]{bruneau13}
L.~Bruneau, V.J.; Pillet, C.A.
\newblock { Landauer-B\"uttiker formula and Schr\"odinger conjecture}.
\newblock {\em { arXiv:1201.3190}} {\bf {2013}}.

\bibitem[Michal~Kolar and Kurizki(2012)]{gershon12}
Kolar, M.; Gelbwaser-Klimovsky, D.; Alicki, R.; Kurizki, G.
\newblock Quantum bath refrigeration towards absolute zero: unattainability
  principle challenged.
\newblock {\em Phys. Rev. Lett.} {\bf 2012}, {\em 108},~090601.

\end{thebibliography}

\end{document}